\documentclass[prl,preprint,showpacs,superscriptaddress,groupedaddress]{revtex4}  
\usepackage{graphicx}  
\usepackage{dcolumn}   
\usepackage{bm}        
\usepackage{amssymb}   
\usepackage{amsmath}
\usepackage[colorlinks=true, citecolor=blue, urlcolor=blue, linkcolor = blue]{hyperref}
\usepackage{bm}
\usepackage{color}
\usepackage{bookmark}
\usepackage{tabularx}
\usepackage{mathtools}
\usepackage{microtype}
\usepackage{cancel}
\usepackage{relsize}
\usepackage{float}
\usepackage[caption = false]{subfig}
\usepackage{xcolor}

\begin{document}

\title{Euclidean Q-balls of fluctuating SDW/CDW in the 'nested' Hubbard model of high-T$_c$ superconductors as the origin of pseudogap and superconducting behaviors}

\affiliation{Theoretical Physics and Quantum Technologies Department, NUST "MISIS", Leninskiy ave. 4, 119049 Moscow, Russia; i.m.sergei.m@gmail.com}
\author{Sergei I. Mukhin}



\date{\today}





\begin{abstract}The origin of the pseudogap (PG) phase and superconducting behaviors in high-Tc superconductors is proposed, based on the picture of Euclidean Q-balls formation, that carry Cooper/local-pair condensates inside their volumes.  Unlike the baryonic Q-balls in supersymmetric standard model, the Euclidean Q-balls describe  spin-/charge densities (SDW/CDW), that oscillate in Matsubara time, and are found in the 'nested' repulsive Hubbard model of high-Tc superconductors. Euclidean Q-balls arise due to global invariance of the effective theory under the phase rotation of the Fourier amplitudes of SDW/CDW fluctuations, leading to conservation of the 'Noether charge' Q in Matsubara time. Due to local minimum of their potential energy at finite amplitude of the density fluctuations, the Q-balls provide greater binding energy of fermions into local/Cooper pairs than one created by exchange with infinitesimal lattice/charge/spin quasiparticles in the usual Fr\"{o}hlich mechanism. We show that below some temperature T$_n^*$ the Q-balls arise with a finite density of superconducting condensate inside them. The Q-balls expand their sizes to infinity at superconducting transition temperature T$_c$.  Fermionic spectral gap inside the Q-balls arises in the vicinity of the 'nested' regions of the bare Fermi surface. Solutions are found analytically from the Eliashberg like equations with the 'nesting' wave vectors connecting different 'hot spots' on the bare Fermi surface. The experimental 'Uemura plot' (1989) of T$_c$ {\it{versus}} superconducting density $n_s$, as well as experimental plots of diamagnetic moment {\it{versus}} magnetic field above Tc in cuprates by  Li et al. (2010), follow naturally from the proposed theory. The 'breathing modes' of the Q-balls in Matsubara time, as well as sharp maximum in specific heat temperature dependence in the vicinity of the 1-st order phase transition into Q-balls gas phase are also predicted.
\end{abstract}

\maketitle








\section{Introduction}

A theory is presented of  'Euclidean Q-ball phase' of high-T$_c$ cuprates in the 'nested' Hubbard model, that may explain both the high-T$_c$ superconductivity, as well as  the 'pseudo gap' phase, that precedes it. A new mechanism of binding the fermions into local/Cooper pairs via exchange with density fluctuations of finite amplitude, the Q-balls, instead of usual Fr\"{o}hlich  mechanism is proposed. The spin/charge fluctuations inside the Q-balls possess local minimum at finite amplitude and therefore provide greater binding energy of fermions into local/Cooper pairs than exchange with infinitesimal lattice/charge/spin quasiparticles in the usual Fr\"{o}hlich picture. In this work it is demonstrated analytically that effective Euclidean action of the oscillating in Matsubara time spin-/charge densities (SDW/CDW), obtained by integrating out self-consistently emergent Cooper/local-pair degrees of freedom, possesses a local minimum at finite amplitude of the oscillations. Euclidean Q-balls arise due to global invariance of the effective theory under the phase rotation of the Fourier amplitudes of SDW/CDW fluctuations, leading to conservation of the 'Noether charge' Q in Matsubara time. This is reminiscent of the famous Q-balls formation in the supersymmetric standard model, where the Noether charge responsible for baryon number conservation is associated with the U(1) symmetry of the squarks field \cite{Coleman, Lee and Pang}. As a result, we found that at some temperature $T^*$ the leading collective spin-/charge fluctuations acquire  the form  of finite volume Q-balls filled with Cooper/local-pair condensates. The dominating Fourier  component of these  spin-/charge density wave fluctuations, that causes local/Cooper pairing, rotates with bosonic Matsubara frequency in the Euclidean space-time, while the local minimum of Q-ball potential energy is located at finite value of the modulus of the Fourier amplitude. Simultaneously, it is demonstrated that the 'gas' of Q-balls  arises at T$^*$ as a 1$^{st}$ order  phase transition. This mechanism of local/Cooper pairing provides the major distinction from the usual phonon- \cite{elis}or spin-fermion coupling models considered for high-T$_c$ cuprates: spin-waves \cite{Chubukov}, charge-density fluctuations \cite{Seibold}, or polarons \cite{Bianconi}, where a `pairing glue' between the paired fermions is provided by exchange with phonons or spin-/charge waves of infinitesimal amplitudes. The superconducting transition happens at T$_c$, where either the Q-balls form infinite percolating cluster for the local/Cooper-pairs, or the minimum of the Q-balls potential energy crosses zero of energies, thus, making the Q-ball volume infinite.   
The plan of the article is as follows. In Section I an effective U(1) symmetric Euclidean model of the SDW/CDW fluctuations described by a scalar amplitude field is outlined and condition for the Q-ball emergence is derived. Section II contains derivation of the effective potential energy of the  SDW/CDW fluctuations, induced by formation of a local superconducting condensate inside the Q-balls.  In Section III the local superconducting 'pseudo gap' inside a Q-ball  is self-consistently derived from the Eliashberg like equation, that acquires form of the Mathieu equation  with the Matsubara time as a coordinate, while the propagator of the semiclassical SDW/CDW fluctuations plays the role of the periodic potential.  Temperatures  T$^*$ and T$_c$  are expressed in analytic form as functions of the spin/charge-fermion coupling constant, density of the  'nested' states,  and short-range coherence length of the spin-/charge density waves in the strongly correlated electron system. In Section IV the experimental 'Uemura plot' \cite{Uemura} of T$_c$ {\it{versus}} superconducting density $n_s$ is compared with the present theory results and qualitative correspondence is found. In Section V the size of a Q-ball and Q-ball space-dependent spherically symmetric solution is found in analytic form. In Section VI the entropy and specific heat of the Q-ball gas is calculated, demonstrating sharp maximum as function of temperature at the 1-st order phase transition into Q-ball phase.The 'breathing modes' of the Q-balls in Matsubara time are found analytically.  In Section VII the Q-ball gas  diamagnetic moment {\it{versus}} magnetic field above Tc is calculated and favourably compares with experimental plots in cuprates by  Li et al. (2010) \cite{li}. The future applications of the presented model for description of the properties of high-T$_c$ cuprates are discussed in the Conclusions.  
 
\section{Effective model}

We consider a simplest model Euclidean action ${S}_{M}$ with a scalar complex field $M(\tau,{\bf{r}})$ related with spin-/charge- density fluctuations :

 \begin{eqnarray}
S_{M}=\int_0^{\beta}\int_Vd\tau d^D{\bf{r}}\dfrac{1}{g}\left\{ |\partial_{\tau}M|^2 +s^2 |\partial_{{\bf{r}}}M|^2 + {\mu _0 ^2 }{|M|}^2+ gU_{f}(|M|^2)\right\},\; M\equiv M(\tau,{\bf{r}})\,, \label{Eu}
\end{eqnarray}
\noindent where $M(\tau+1/T,{\bf{r}})=M(\tau,{\bf{r}})$ is periodic function of Matsubara time at finite temperature $T$ \cite{agd}, $s$ is bare propagation velocity, correlation length $\xi$ of the fluctuations is defined by the 'mass' term $\mu_0^2\sim 1/\xi^2$, and effective potential $U_{f}(|M|^2)$ depends on the field modulus $|M|$ and contains charge-/spin-fermion coupling constant $g$ in front. In what follows, Eq. (\ref{Eu}) is used to describe effective theory of  the Fourier components of  the leading SDW/CDW fluctuations. Explicit expression for $U_{f}(|M(\tau,{\bf{r}})|)$ is derived below by integrating out Cooper/local-pairs fluctuations in the 'nested' Hubbard model with charge-/spin-fermion interactions. The model (\ref{Eu}) is $U(1)$ invariant under the global phase rotation $\phi$: $M\rightarrow Me^{i\phi}$. Hence, corresponding  'Noether charge' is conserved along the Matsubara time axis. The 'Noether charge' conservation makes possible Matsubara time periodic, finite volume Q-ball semiclassical solutions, that otherwise would be banned in D$>2$ by Derrick theorem \cite{Derrick} in the static case. Previously Q-balls were introduced by Coleman \cite{Coleman} for Minkowski space in QCD, and have been classified as non-topological solitons \cite{Lee and Pang}. As shown below, the Euclidean Q-balls  describe stable semiclassical short-range charge/spin ordering fluctuations of finite energy, that appear at finite temperatures below some temperature T$^*$ found below. The fermionic spectral gap inside Euclidean Q-balls arises in the vicinity of the 'nested' regions of the bare Fermi surface (corresponding to the antinodal points of the cuprates Fermi-surface) and scales with the local superconducting density inside Q-balls. Hence, $T^{*}$ defines temperature of a phase transition into the 'pseudo gap' phase, as was proposed previously \cite{Mukhin}. 

Consider now time-dependent phase shift: $\phi=\Omega \tau$, with frequency $\Omega=2\pi n T$, that satisfies Matsubara time periodicity. Then, the corresponding conserved  'Noether charge' is found readily. First, one defines $D+1$-dimensional 'current density' $\{j_{\tau},\vec{j}\}$ of the scalar field $M(\tau,{\bf{r}})$:

 \begin{eqnarray}
j_{\alpha}=\frac{i}{2}\left\{M^{*}(\tau,{\bf{r}})\partial_\alpha M(\tau,{\bf{r}})-M(\tau,{\bf{r}})\partial_\alpha M^{*}(\tau,{\bf{r}})\right\}, \;\alpha=\tau,{\bf{r}}.\label{J}
\end{eqnarray}
\noindent Next, Euclidean trajectories of the field, defined by 'classical' equations of motion are considered:

 \begin{eqnarray}
\dfrac{\delta S_{M}}{\delta M^*(\tau,{\bf{r}})}=-\partial^2_\tau M(\tau,{\bf{r}})-s^2\sum_{\alpha={\bf{r}}}\partial^2_\alpha M(\tau,{\bf{r}})+\mu _0 ^2 {M(\tau,{\bf{r}})}+gM(\tau,{\bf{r}})\dfrac{\partial U_f}{\partial 
|M(\tau,{\bf{r}})|^2 }=0.\label{L}
\end{eqnarray}
\noindent Using Eqs. (\ref{J}), (\ref{L}) it is straightforward to prove the following relation:

\begin{eqnarray}
\dfrac{\partial}{\partial\tau}\int_V j_{\tau}d^D{\bf{r}}=-s^2\int_V div \vec{j}\,d^D{\bf{r}}=-s^2\oint_{S(V)}\vec{j}\cdot d\vec{S}\;,\label{N}
\end{eqnarray}
\noindent where the last integral in Eq. (\ref{N}) is taken over the surface $S$ of the volume $V$ due to the Gauss theorem. Hence, for the non-topological field configurations, that occupy finite volume $V$ ,  i.e. $M(\tau,{\bf{r}}\notin V)\equiv 0$, one finds:

\begin{eqnarray}
\dfrac{\partial}{\partial\tau}\int_V j_{\tau}d^D{\bf{r}}=-s^2\oint_{S(V)}\vec{j}\cdot d\vec{S}=0\;,\label{NN}
\end{eqnarray}
\noindent and, in turn, conserved 'Noether charge' $Q$ equals:

\begin{eqnarray}
Q=\int_V j_{\tau}d^D{\bf{r}}= \Omega M(\tau)^2V\; , \label{Q}
\end{eqnarray}
\noindent where we have approximated the 'Q-ball' field configuration with a step function $\Theta ({\bf{r}})$: 

\begin{align}
M(\tau,{\bf{r}})=e^{-i\Omega\tau}M\Theta\left\{{\bf{r}}\right\}\;;\quad \Theta({\bf{r}})\equiv\begin{cases}1 ;\;{\bf{r}}\in V ;\\
0;\;{\bf{r}}\notin V.\end{cases}
\label{step}
\end{align}
\noindent
In general, to find equilibrium volume of the Q-ball one has to minimise the action $S_{M}$ with respect to $V$ under the conserved  'charge' Q defined by Eq. (\ref{Q}). First we do this in the step function approximation above, Eq. (ref{step}). In this case
one finds action $S_{M}$ using Eqs. (\ref{Eu}) and (\ref{Q}), and neglecting the boundary contribution $\propto \int|\partial_{{\bf{r}}}M(\tau,{\bf{r}})|^2 $:

 \begin{eqnarray}
 S_{M}=V\frac{1}{gT }\left\{[\Omega^2 +{\mu _0 ^2 }]M^2+ gU_{f}\right\}=\frac{1}{gT }\left\{\dfrac{Q^2}{VM^2}+ V[\mu _0 ^2M^2 + gU_{f}]\right\}. \label{SMQ}
\end{eqnarray}
\noindent Minimising Euclidian action of the Q-ball in Eq. (\ref{SMQ}) with respect to volume $V$ one finds:

\begin{eqnarray}
 \dfrac{\partial S_{M}}{\partial V}=\frac{1}{gT}\left\{-\dfrac{Q^2}{V^2M^2}+[\mu _0 ^2 M^2+ gU_{f}(M^2)]\right\}=0. \label{min}
\end{eqnarray}
\noindent Solving Eq. (\ref{min}) one finds equilibrium volume $V_Q$ of the Q-ball and its energy $E_Q$:

\begin{eqnarray}
 V_Q= \dfrac{Q}{M\sqrt{\mu _0 ^2 M^2+ gU_{f}(M^2)}}\;.\; \label{VQ}
 \end{eqnarray}
 \noindent Substituting Eq. (\ref{VQ}) into Eq. (\ref{SMQ}) one finds:
\begin{eqnarray}
 E_Q=TS^{min}_M=\dfrac{2Q\sqrt{\mu _0 ^2 M^2+ gU_{f}(M^2)}}{gM}=\dfrac{2Q\Omega}{g}, \label{aQ}
\end{eqnarray}
\noindent where the last equality follows directly after substitution of  expression $V_Q$ from Eq. (\ref{VQ}) into Eq. (\ref{Q}). 
Since $Q$ cancels in Eq. (\ref{aQ}), the following self-consistency equation follows:  

\begin{eqnarray}
 \Omega=\dfrac{\sqrt{\mu _0 ^2 M^2+ gU_{f}(M^2)}}{M}. \label{self}
\end{eqnarray}
\noindent  In a more careful procedure, that uses Eq. (\ref{L}) to extract exact coordinate dependence of $M(\tau,{\bf{r}})$ one has to add to the action a term with the Lagrange multiplier, that takes care for the 'charge' Q conservation:

 \begin{eqnarray}
&S_{M}=\int_0^{\beta}\int_Vd\tau d^D{\bf{r}}\dfrac{1}{g}\left\{ |\partial_{\tau}M|^2 +s^2 |\partial_{{\bf{r}}}M|^2 + {\mu _0 ^2 }{|M|}^2+ gU_{f}(|M|^2)\right.+\nonumber \\
&\left. i\mu\left\{M^{*}(\tau,{\bf{r}})\partial_\alpha M(\tau,{\bf{r}})-M(\tau,{\bf{r}})\partial_\alpha M^{*}(\tau,{\bf{r}})\right\}\right\},\; M\equiv M(\tau,{\bf{r}})\, \label{Eum}
\end{eqnarray}
\noindent It is easy to find, that the value of the 'chemical potential' $\mu$ should be $\mu=-\Omega$, in order to recover from Eq. (\ref{Eum}) the approximate self-consistency equation  Eq. (\ref{self}) in the step-function approximation Eq. (\ref{step}). Then, substituting $S_{M}$ from Eq. (\ref{Eum}) into dynamic equation Eq. (\ref{L}) in Euclidean space-time and using for the time-dependence $M(\tau,{\bf{r}})\propto \exp\{-i\Omega t\}$ one finally obtains the coordinate dependent self consistency equation, to be solved below:

\begin{eqnarray}
-s^2\Delta M+\partial U_f/\partial M+(\mu_0^2-\Omega^2)M=0, \label{selfr}
\end{eqnarray}
\noindent compare \cite{Coleman}. 
An Euclidean Q-ball described by Eqs. (\ref{Q}), (\ref{self}) and (\ref{selfr}) differs from the Q-ball in Minkowski space \cite{Coleman}: at fixed temperature $T$ a choice of the values of the Matsubara frequencies $\Omega=2\pi nT$ in Euclidean space-time is discrete due to integer $n$ and starts from $\Omega=2\pi T$, contrary to a continuum of the frequency values in the Minkowski space-time. Hence, the highest temperature T$^*$, at which  Eq. (\ref{self}) possesses solution, would be for $n=1$, and would manifest transition into Q-ball 'gas' phase, corresponding to a PG phase, as will be shown below. Next, at temperature T$_c <$ T$^*$, the Q-ball energy becomes zero,  $E_Q=0$, in Eq. (\ref{aQ}):
\begin{eqnarray}
 0=\dfrac{\sqrt{\mu _0 ^2 M^2+ gU_{f}(M^2)}}{M}. \label{sc}
\end{eqnarray}
\noindent Then, Q-ball volume becomes infinite in accord with Eq. (\ref{VQ}), and a phase transition into bulk superconducting phase takes place. One has to derive an explicit expression for the effective energy $U_f(M^2)$ in order to explore the phase diagram of the Q-balls 'gas' in the next Sections.
\section{Free energy of the Cooper-pairing fluctuations inside the Q-balls}

\begin{figure}[h!!]
\centerline{\includegraphics[width=0.5\linewidth]{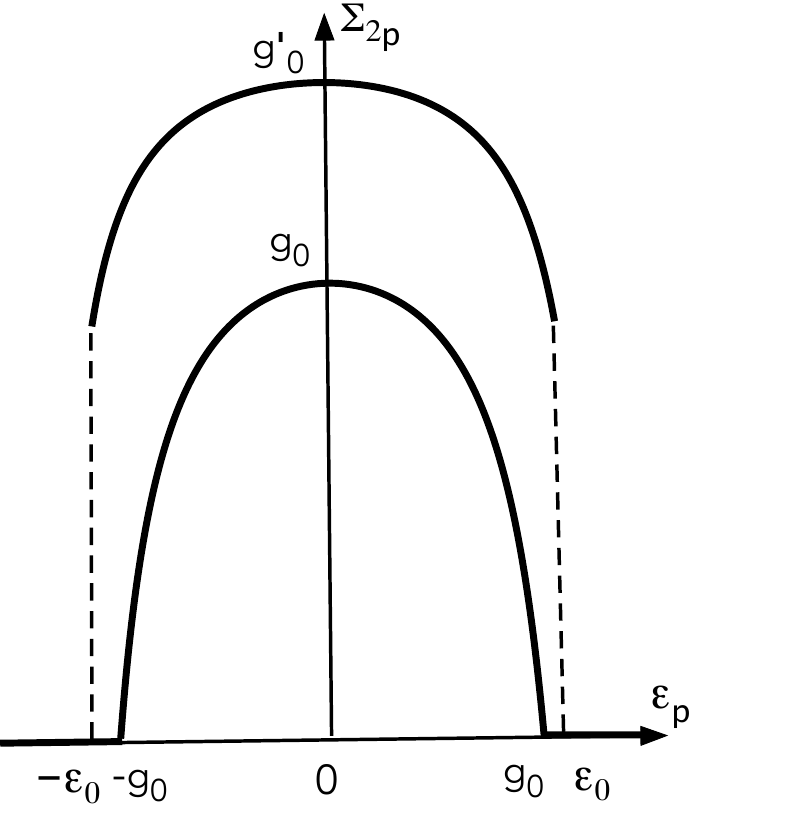}}
\caption{ Anomalous self-energy $\Sigma_{2p,\sigma}$ (fermionic spectral gap) as function of fermionic dispersion $\varepsilon_p$ in the vicinity of the Fermi-level $\mu=0$ near the 'nested'/anti-nodal  points  of the bare Fermi surface inside a Q-ball, for two cases of the local superconducting PG $g_0 < \varepsilon_0$ and $g'_0 > \varepsilon_0$, see text.}
\label{1}
\end{figure}

Here we derive an effective potential $U_{f}(|M({\bf{r}})|)$, being the density of the free energy decrease $\Delta\Omega_s$ due to superconducting fluctuations. Consider a simple model of fermions on a square lattice, that are linearly coupled to the dominant Q-ball type charge- or spin density fluctuations, that obey Eq. (\ref{L}), and possess amplitude $M(\tau,{\bf{r}})\equiv e^{-i\Omega\tau}M({\bf{r}}) $ with wave vectors $Q_{CDW}$ or $Q_{SDW}$ respectively. In what follows we accept generalised notation $Q_{DW}$ for both cases. Thus, the fermionic part of the Euclidean action $S_{f}$ takes the form: 

 \begin{eqnarray}
{S}_{f}=\int_0^{\beta}d\tau\int_Vd^D{\bf{r}}\sum_{{\bf{q}},\sigma}\left[ c^{+}_{{\bf{q}}\sigma}(\partial_\tau+\varepsilon_q)c_{{\bf{q}},\sigma}+\left(c^{+}_{{\bf{q+Q_{DW}}},\sigma}M(\tau,{\bf{r}})\sigma c_{{\bf{q}},\sigma}+H.c.\right)\right]\;, \label{f}
\end{eqnarray}

\noindent and antiferromagnetic fluctuations are considered below for definiteness using standard Hamiltonian \cite{Chubukov} with spin-fermion coupling. Then, the Matsubara time periodic complex amplitude $M(\tau,{\bf{r}})$ considered in general in the preceding section, acquires a particular meaning of the amplitude of the SDW fluctuation, with the fast space oscillations on the lattice variable ${\bf{r}}$ being characterised by a wave-vector $Q_{DW}$, and slow variations on the scale of the correlation length or Q-ball radius:
\begin{eqnarray}
&&{M}_{\tau,{\bf{r}}}={M}(\tau,{\bf{r}})e^{i{\bf{Q_{DW}\cdot r}}}+{M(\tau,{\bf{r}})}^{*}e^{-i{\bf{Q_{DW}\cdot r}}},\; \nonumber\\
&&M(\tau,{\bf{r}})\equiv |M(\tau,{\bf{r}})|e^{-i\Omega\tau},\;\Omega=2\pi nT,\; n=1,2,...
\label{SDWQ0}
\end{eqnarray}
\noindent Here $\Omega$ is bosonic Matsubara frequency, and $\sigma$ in Eq. (\ref{f}) is local $z$-axis projection of the fermionic spin assumed to be collinear with the direction of the  spin density inside a Q-ball. A 'slow' $\tau$-dependence of an amplitude $|M(\tau,{\bf{r}})|$ may arise as shown below. Effective theory is then obtained by formally integrating out fermions, assuming that they undergo local Cooper/local-pairing fluctuations with emerging Bogoliubov anomalous averages $<c_{{\bf{q}},\sigma}c_{-{\bf{q}},-\sigma}>$, $<c^{+}_{{\bf{q}},\sigma}c^{+}_{-{\bf{q}},-\sigma}>$ entering the diagrammatic expansion of the free energy $\Omega_s$ \cite{Mukhin}:

\begin{align}
&VU_{f}(|M(\tau,{\bf{r}})|)=\Delta\Omega_s=-T\ln\dfrac{Tr\left\{e^{-\int_0^{\beta}H_{int}(\tau)d\tau}{\cal{G}}(0)\right\}}{Tr\left\{{\cal{G}}(0)\right\}}\equiv \Omega_s-\Omega_0;\,{\cal{G}}(0)\equiv e^{-\beta H_0};\,\label{DOI} \\
&H_{int}=\int_Vd^D{\bf{r}}\sum_{{\bf{q}},\sigma}\left(c^{+}_{{\bf{q+Q_{DW}}},\sigma}M(\tau,{\bf{r}})\sigma c_{{\bf{q}},\sigma}+H.c.\right),
\label{intinst0}
\end{align}

\noindent where $H_0$ is inferred from the first and $H_{int}$ from the second term in the sum in (\ref{f}) respectively. Next, we multiply Hamiltonian $H_{int}$ in (\ref{intinst0}) with a dimensionless  amplitude $0<\alpha<1$, as a formal variable coupling strength in the spin-fermion interaction, and calculate the free energy derivative in accord with the usual prescription \cite{agd}:
\begin{align}
&\dfrac{\partial\Omega_s}{\partial \alpha}=T{\int_0^{\beta}{\left\langle\dfrac{\partial H_{int}(\tau)}{\partial \alpha}\right\rangle} d\tau}=
-\frac{T}{\alpha}\int_0^{\beta}{\int}_0^{\beta}d\tau d\tau_1 \left\langle H_{int}(\tau) H_{int}(\tau_1)\right\rangle =
\nonumber\\
&-\frac{TV}{\alpha}|M|^2T\sum_{\omega,{\bf{p}},\sigma}\sigma\bar{\sigma}\overline{F}_{\sigma,\bar{\sigma}}(\omega,{\bf{p}}){F}_{\bar{\sigma},\sigma}(\omega-\Omega,{\bf{p}}-{\bf{Q_{DW}}})\alpha^2 \;,
\label{DOs}
\end{align}

\noindent where we have neglected slow dependence of the modulus of the SDW amplitude $|M|$ on $\tau, {\bf{r}}$ in the step function approximation (\ref{step}). The loop of Gor'kov anomalous functions $F^\dagger, F$ connected with the 'gluon' line $D(\tau-\tau')\sim {M(\tau')}^*\cdot M(\tau)$, depends now on parameter $\alpha$. The amplitudes $M$ in Eq. (\ref{SDWQ0}) of spin-/charge density fluctuations obey  'classical' equations of motion Eq. (\ref{L}), that extremize Euclidean action. In the case when wave vector $\bf{Q_{DW}}$ connects 'nested' points on the Fermi surface belonging to the regions with opposite signs of the d-wave superconducting order parameter, the following algebraic relations hold for the dispersion and self-energy functions \cite{Mukhin}: 

\begin{align}
&\varepsilon _{p - Q_{DW}}  =  - \varepsilon _p\equiv -\varepsilon ;\quad \Sigma _{2p - Q_{DW},\sigma }  =  - \Sigma _{2p,\sigma };\quad {\Sigma}^*_{1p,\sigma}(\omega)\equiv\Sigma_{1,-p,\overline{\sigma}}(-\omega)\;;\label{dnes}\\
&F_{p,\sigma}(\omega) = \frac{-\Sigma_{2p,\sigma}}
{|  i\omega - \varepsilon _{p }  - \Sigma _{1p ,\sigma } (\omega)|^2+|\Sigma_{2p,\sigma}(\omega)|^2},\quad \omega=\pi (2n+1)T;\quad n=0,\pm1,...\label{F}
\end{align}
\noindent In what follows we neglect renormalisations \cite{Mukhin} entering via self-energy $\Sigma_{1p,\sigma}(\varepsilon,\omega)$ in denominator in Eq. (\ref{F}) for the anomalous fermionic Green function $F_{p,\sigma}(\omega)$, and use $d$-wave symmetric behaviour of superconducting order parameter $\Sigma _{2p - Q_{DW},\sigma }  =  - \Sigma _{2p,\sigma }$ represented by the self-energy function $\Sigma _{2p,\sigma }$. The latter is approximated with parabolic function of bare fermionic dispersion $\varepsilon_p$ in the vicinity of the Fermi energy, see Eq. (\ref{freeint}) below. Now, substituting expressions in Eqs. (\ref{dnes}) and (\ref{F}) into Eq. (\ref{DOs}) one finds:

\begin{align}
\dfrac{\partial\Omega_s}{\partial \alpha}= -{TV}{R\alpha};\quad R=\dfrac{4M^2\nu g_0\varepsilon_0}{3T(\Omega^2+4g_0^2)}\tanh{\dfrac{g_{0}}{2T}}\tanh{\dfrac{g_{0}}{\varepsilon_0}};\quad g_0^2\equiv \varepsilon_p^2+|\Sigma_{2p,\sigma}(\omega)|^2\;.
\label{freeint}
\end{align}
\noindent Here expression for $R$ in Eq. (\ref{freeint}) is obtained after summation over fermionic frequency $\omega=\pi (2n+1)$ in Eq. (\ref{DOs}), while neglecting $\omega$-dependence of the self-energy $\Sigma_{2p,\sigma}(\omega)\approx \Sigma_{2p,\sigma}(0)$, since summation in Eq. (\ref{DOs}) over $\omega$ is quickly convergent. Summation over momenta ${\bf{p}}$ in Eq. (\ref{DOs}) is substituted by integration over $\varepsilon_p$ (counted from the Fermi level $\mu$) over bare density of 'nested' states $\nu(\varepsilon_p)$ approximated as:

\begin{align}
\nu(\varepsilon_p)=\begin{cases}\nu ;\;|\varepsilon_p|\leq \varepsilon_0;\\
0;\;|\varepsilon_p| > \varepsilon_0.\end{cases}
\label{nunest}
\end{align}
\noindent Simultaneously, $|\Sigma_{2p,\sigma}(\omega)|^2=g_0^2-\varepsilon_p^2\geq 0$ differs from zero inside an interval: $-g_0\leq\varepsilon_p\leq g_0$, see Fig.\ref{1}. Hence, in Eq. (\ref{freeint}) product $\nu g_0\varepsilon_0\tanh{{g_{0}}/{\varepsilon_0}}$ interpolates between the cases $g_0 > \varepsilon_0$ and $g_0 < \varepsilon_0$.

\noindent
Now, one has to bear in mind that $g_0=g_0(\alpha)$, and, hence, $R(\alpha)$ defined in Eq. (\ref{freeint}), depends on the integration variable $\alpha$ introduced above. To complete derivation of the effective potential $U_{f}(|M(\tau,{\bf{r}})|)$ one has to find constant $g_0^2$ entering expression for $R$.  The local 'superconducting PG' $g_0$ is found from the self-consistency condition derived below, see also \cite{Mukhin}. Importantly, the final expression of the kind obtained in Eq. (\ref{freeint}) appears also in the case when charge fluctuations instead of spin fluctuations couple to the fermions via interaction Hamiltonian:

\begin{align}
H^{CDW}_{int}=\int_Vd^D{\bf{r}}\sum_{{\bf{q}},\sigma}\left(c^{+}_{{\bf{q+Q_{DW}}},\sigma}M_{C}(\tau,{\bf{r}}) c_{{\bf{q}},\sigma}+H.c.\right),
\label{intinstC}
\end{align}
\noindent where $\sigma$ spin factor is missing in the charge - fermion coupling vertex $c^{+}M_Cc$. This would, in turn, lead to the absence of the factor $\sigma\bar{\sigma}=-1$ in the Eq. (\ref{DOs}). Hence, in order to keep $U_f <0$ (the driving force of the Q-ball transition) one has to compensate for this sign change. For this, it is necessary to change the sign of the Green's functions product $\overline{F}_{\sigma,\bar{\sigma}}(\omega,{\bf{p}}){F}_{\bar{\sigma},\sigma}(\omega-\Omega,{\bf{p}}-{\bf{Q_{DW}}})$ in Eq. (\ref{DOs}). Then, allowing for the structure of the Gor'kov's anomalous Green's function in Eq. (\ref{F}) one concludes, that relation between the superconducting order parameters in the points connected by the 'nesting' wave vector $Q_{CDW}$  should be altered , i.e in case of CDW-mediated pairing the 'nesting' wave vector should couple points with the same sign of superconducting order parameter: $\Sigma _{2p - Q_{CDW},\sigma }  =  \Sigma _{2p,\sigma }$.

\section{Eliashberg Equations and Bound States Along the Axis of Matsubara Time}

Now, using definition of the anomalous fermionic Green's function $F_{p,\sigma}(\omega)$ in Eq. (\ref{F}), one obtains Eliashberg equation for the self-energy $\Sigma_{2p,\sigma}(\omega)$ \cite{elis, Mukhin} in the form:

\begin{eqnarray}
\Sigma _{2p,\sigma } (\omega ) = \displaystyle \frac{ - T{\cal{D}}_{Q_{DW}} (\Omega )\Sigma _{2,p - Q_{DW},\sigma } (\omega  - \Omega )}{|  i(\omega  - \Omega ) - \varepsilon _{p - Q_{DW}}  - \Sigma _{1p - Q_{DW},\sigma } (\omega  - \Omega )|^2  + |\Sigma _{2p - Q_{DW},\sigma } (\omega  - \Omega )|^2 }\,,
\label{sigs20}\\
{\cal{D}}_{Q_{DW}} (\Omega )\equiv \dfrac{M^2}{T} \,, \label{DM}
\end{eqnarray}
\noindent where expression Eq. (\ref{DM}) for the 'glue boson' propagator is inferred from the definition of the considered above 'classical' Q-ball field $M(\tau,{\bf{r}})$ , as defined in Eqs. (\ref{step}) and (\ref{SDWQ0}), and monochromaticity of the 'glue boson' propagator is taken into account, thus, transforming equation (\ref{sigs20}) into algebraic.  It is easy to compare Eqs. (\ref{sigs20}) and (\ref{F}) and obtain readily equation for the anomalous Green function $F_{p,\sigma}(\omega)$ in the closed form (compare \cite{Mukhin}):

\begin{eqnarray}
F_{p,\sigma}(\omega)=-\Sigma _{2p,\sigma } (\omega )K_{p}(\omega)=K_{p}(\omega)\left[{{\cal{D}}_{Q_{DW}} (\Omega )F_{p-Q_{DW},\sigma}(\omega  - \Omega )}\right]\;,\label{Feq} \\
K_{p}(\omega)=\left\{|  i\omega - \varepsilon _{p }  - \Sigma _{1p ,\sigma } (\omega)|^2+|\Sigma_{2p,\sigma}(\omega)|^2\right\}^{-1}\approx \left\{\omega^2 + \varepsilon _{p }^2 +|\Sigma_{2p,\sigma}(\omega)|^2\right\}^{-1}. \label{K0}\end{eqnarray}

\noindent When writing Eq. (\ref{Feq}) the d-wave symmetry of the self-energy: $\Sigma _{2p - Q_{DW},\sigma }  =  - \Sigma _{2p,\sigma }$, was taken into account. Now, after applying inverse Fourier transform to both sides of Eq. (\ref{Feq}) one finds:

\begin{equation}
F_{p,\sigma} \left( {\tau} \right) = \int_0^{{1 \mathord{\left/{\vphantom {1 T}} \right. \kern-\nulldelimiterspace} T}} {K_{p}\left( {\tau - \tau '} \right)} {\cal{D}}_{Q_{DW}}\left( \tau '  \right)F_{p,\sigma}\left( \tau ' \right)d\tau '\;.\label{intsgm}
\end{equation}
\noindent  Approximating denominator of $K_p(\omega)$ as indicated in Eq. (\ref{K0}), one finds:
\begin{align}
K_{p}(\tau)=T\sum_\omega\dfrac{e^{-i\omega t}}{|  i\omega - \varepsilon _{p }  - \Sigma _{1p ,\sigma } (\omega)|^2+|\Sigma_{2p,\sigma}(\omega)|^2}\approx\frac{{\text{sinh}\left[ {g_0\left( {\frac{1}{{2T}} - \left| \tau  \right|} \right)} \right]}}{{2g_0 \text{cosh}\left( {\frac{g_0}{{2T}}} \right)}}\;,\label{Ktu} 
\end{align}
\noindent where $g_0$ is defined in Eq. (\ref{freeint}). It is straightforward to check that (\ref{Ktu}) possesses the following property:
\begin{equation}
\partial^2_{\tau}K(\tau)=g_0^2K(\tau)-\delta(\tau), 
\label{Ktdif}
\end{equation} 
\noindent Hence, using the above relation (\ref{Ktdif}) and differentiating Equation (\ref{intsgm}) twice over $\tau$ we obtain the following Schr\"{o}dinger like equation for the wave function $F_{p,\sigma}(\tau)$ of the local/Cooper pair along the Matsubara time axis $\tau$: 
\begin{equation}
-\partial^2_{\tau}F_{p,\sigma}\left( \tau  \right) - {\cal{D}}_{Q_{DW}}\left( \tau  \right)F_{p,\sigma} \left( \tau  \right) =  - g_0^2F_{p,\sigma} \left( \tau  \right).
\label{shrod}
\end{equation}
\noindent Using now expression Eq. (\ref{DM}) for the 'glue boson' propagator ${\cal{D}}_{Q_{DW}}$ one finds, that Gor'kov's anomalous Green function $F_{p,\sigma} \left( \tau  \right)$ of the superconducting condensate inside the Q-ball obeys Mathieu equation \cite{Witt}:

\begin{equation}
\partial^2_{\tau}F_{p,\sigma}\left( \tau  \right) +\left(2M^2\cos{\left( \Omega\tau  \right)} - g_0^2\right)F_{p,\sigma} \left( \tau  \right)=0,\;\quad F_{p,\sigma} \left( {\tau}+\dfrac{1}{T} \right)=-F_{p,\sigma} \left( {\tau} \right)\;, 
\label{Mathieu}
\end{equation}
\noindent where the anti-periodicity condition of the fermionic Green function $F_{p,\sigma} \left( {\tau} \right)$ \cite{agd} is explicitly indicated. Since $\Omega=2\pi nT$ in (\ref{Mathieu}) is bosonic Matsubara frequency, the anti-periodicity condition in Eq. (\ref{Mathieu}) imposes a self-consistency relation between the SDW amplitude $M$ and the 'superconducting PG' $g_0$, that is necessary condition for existence of solution $F_{p,\sigma}\left( \tau  \right)$. To find this self-consistency relation in approximate analytic form one may consider Eq. (\ref{Mathieu}) as Schr\"{o}dinger equation  and approximate potential $V(\tau)=-2M^2\cos{\left( \Omega\tau  \right)}$ with rectangular potential of the amplitude $2M^2$ in the interval $-1/2T\leq\tau\leq 1/2T$,  looking for the odd bound state inside this potential well. Then, it is known that such potential well contains the second lowest possible eigenvalue $-g_0^2$ just crossing zero of energy under the condition \cite{Flug}:  
\begin{equation}
g_0^2 \approx 2M\left(M-\Omega\right)\rightarrow g_0^2(\alpha)\approx 2M\alpha\left(M\alpha-\Omega\right)\;,
\label{Mg0}
\end{equation}
\noindent where at the last step an amplitude $M$ is substituted with $\alpha M$ according to the definition of the formal integration parameter in Eq. (\ref{DOs}). Then, after substitution of  solution Eq. (\ref{Mg0}) into Eq. (\ref{freeint}) one finds the following expression for the function $R(\alpha)$:

\begin{align}
\alpha^2R(\alpha)=\dfrac{4M^2\nu\varepsilon_0 \alpha^2\sqrt{2\alpha M\left(\alpha M-\Omega\right)}}{3T(\Omega^2+8\alpha M\left(\alpha M-\Omega\right))}\tanh{\dfrac{\sqrt{2\alpha M\left(\alpha M-\Omega\right)}}{\varepsilon_0}}\tanh{\dfrac{\sqrt{2\alpha M\left(\alpha M-\Omega\right)}}{2T}}.
\label{UFIN}
\end{align}

\begin{figure}[h!!]
{\includegraphics[width=0.55\linewidth]{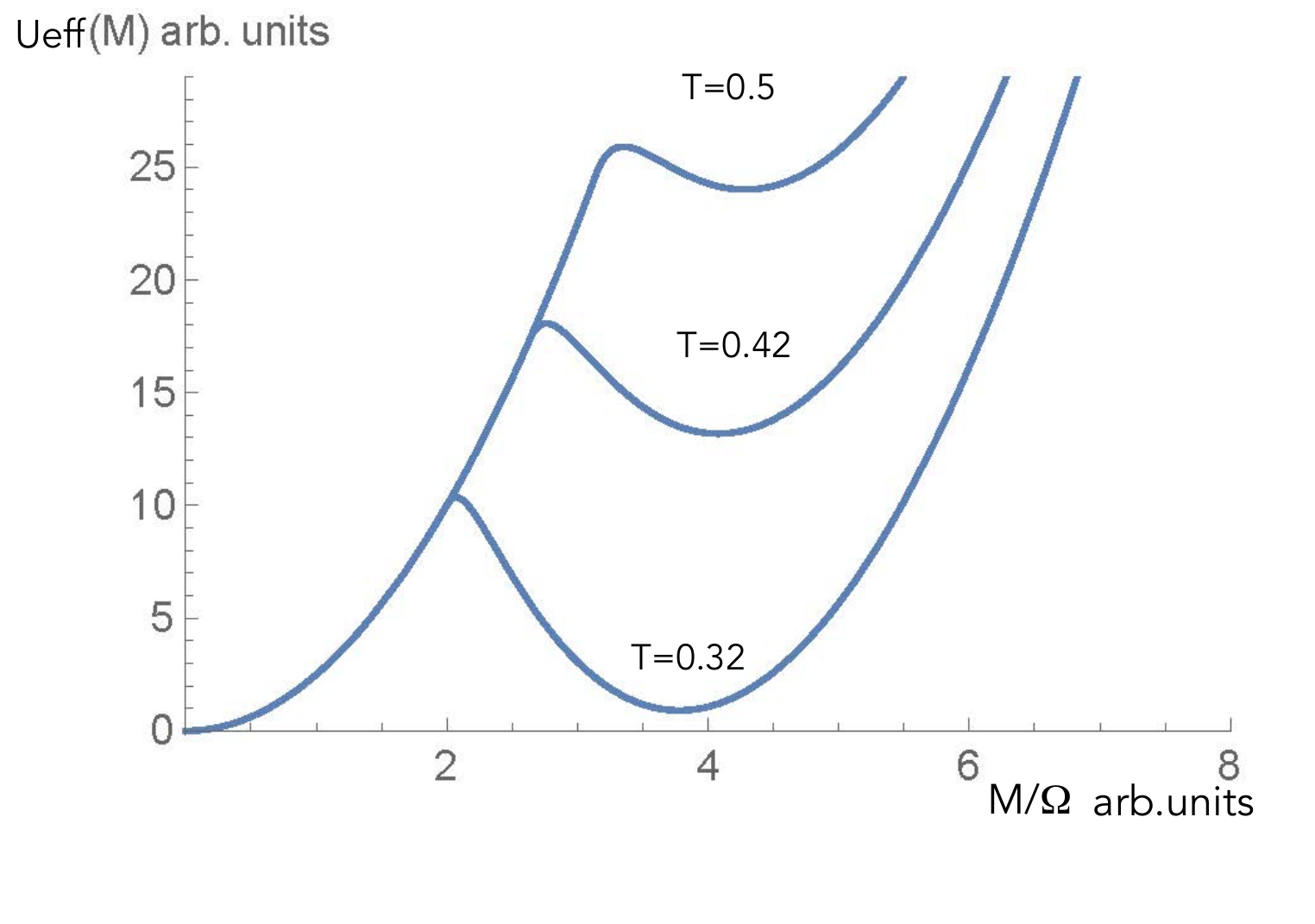}}
\caption{Effective potential energy $U_{eff}$ of the spin/charge density fluctuation for 'mass' $\mu_0=2.23$ and coupling constant  $g\nu\varepsilon_0=59$ in arbitrary units inside a Q-ball as a function of the amplitude $M/\Omega$ weighted by Matsubara frequency $\Omega=2\pi n T$ at three different temperatures $T$ at $n=1$  indicated above the curves, see Eq. (\ref{UFO1}).}
\label{2a}
\end{figure}

\noindent Now, using Eq. (\ref{freeint}) one obtains the following expression for the pairing-induced effective potential energy of SDW/CDW field, that enters Q-ball self-consistency condition in Eq. (\ref{self}):

\begin{align}
&U_{eff}(M)=\mu _0 ^2 M^2+gU_f=\mu _0 ^2 M^2 -\dfrac{4g\nu \varepsilon_0\Omega}{3}I\left(\dfrac{M}{\Omega}\right)\;,\;M\equiv |M(\tau)|\label{UFO1}\\
&I\left(\dfrac{M}{\Omega}\right)= \int_1^{M/\Omega}d\alpha\dfrac{\alpha \sqrt{2\alpha \left(\alpha -1\right)}}{(1+8\alpha \left(\alpha -1\right))}\tanh{\dfrac{\sqrt{2\alpha\left(\alpha-1\right)}\Omega}{\varepsilon_0}}\tanh{\dfrac{\sqrt{2\alpha\left(\alpha-1\right)}\Omega}{2T}}.
\label{UFOI}
\end{align}

\noindent Fig. \ref{2a} contains plots of $U_{eff}(M)$ at different temperatures, manifesting characteristic 'Q-ball local minimum' \cite{Coleman}: near  $ T^*$ temperature, where Q-ball phase has emerged, and  close to $T_c$, at which Q-ball volume becomes infinite and bulk superconductivity sets in.  

Then, it is straightforward to substitute $U_{eff}(M)$ from Eq. (\ref{UFO1}) into self-consistency equation  Eq. (\ref{self}) rewritten by means of  'shifted' by $-M^2\Omega^2$ potential energy $U_{eff}$:
\begin{eqnarray}
\tilde{U}_{eff}\equiv (\mu _0 ^2-\Omega^2){M}^2 - \dfrac{4\Omega \nu \varepsilon_0}{3}I\left(\dfrac{M}{\Omega}\right)=0. \label{self1}
\end{eqnarray}

\begin{figure}[h!!]
{\includegraphics[width=0.45\linewidth]{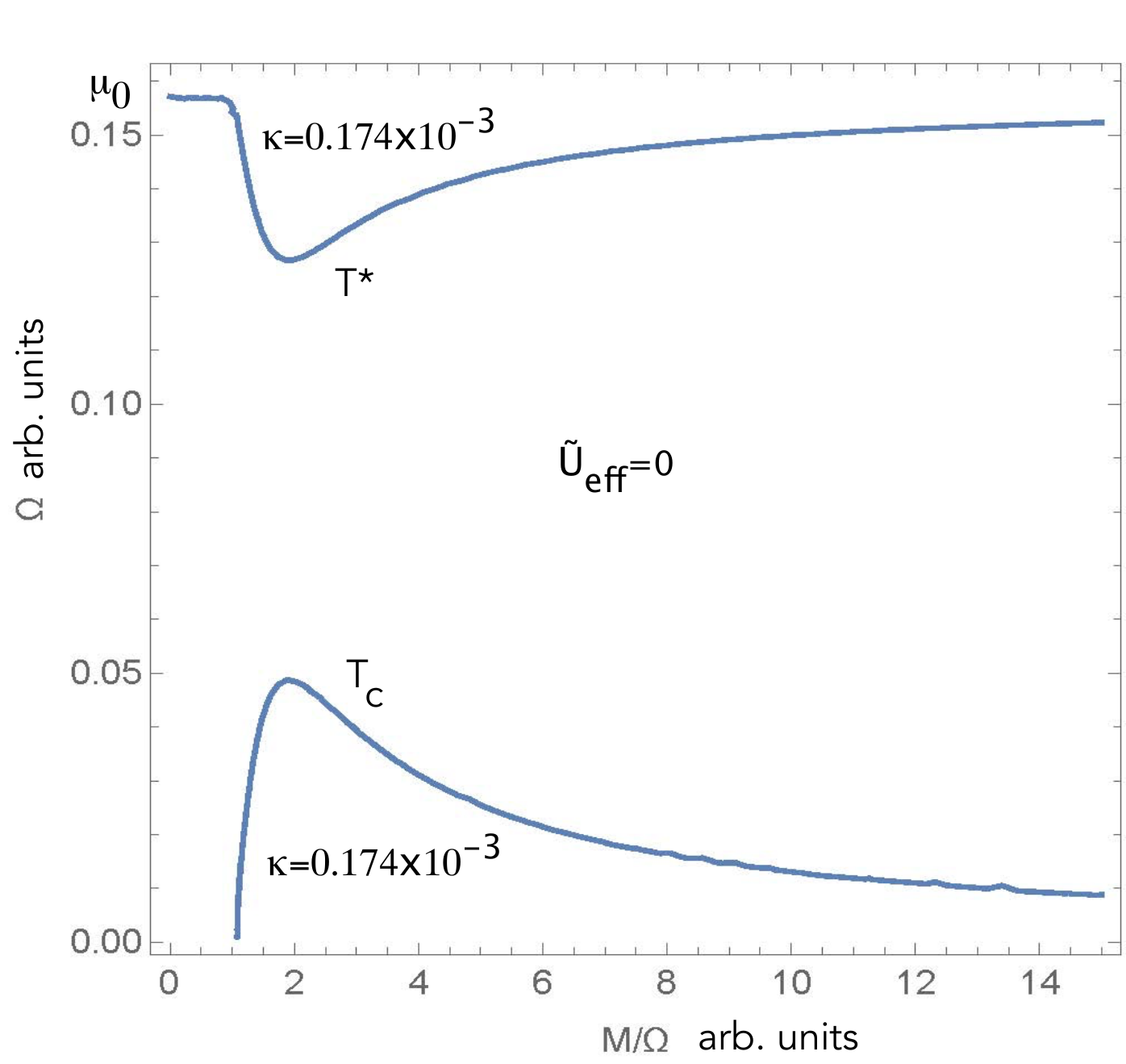}}\vspace{0.1cm}
\caption{The contour plots of self-consistency equation (\ref{self1}) in the plane $\{M/\Omega,\, \Omega\}$ are presented for 'mass' $\mu_0=0.157$ and coupling constant $\kappa\equiv c4g\nu \varepsilon_0/3= 0.174 \cdot 10^{-3}$, in arbitrary units, see text.}
\label{triple}
\end{figure}

 \noindent The contour plots of Eq. (\ref{self1}) in the plane $\left\{M/\Omega,\, \Omega\right\}$ are represented in Fig. \ref{triple} for different ranges of the coupling strength. 

\begin{figure}[h!!]
{\includegraphics[width=0.55\linewidth]{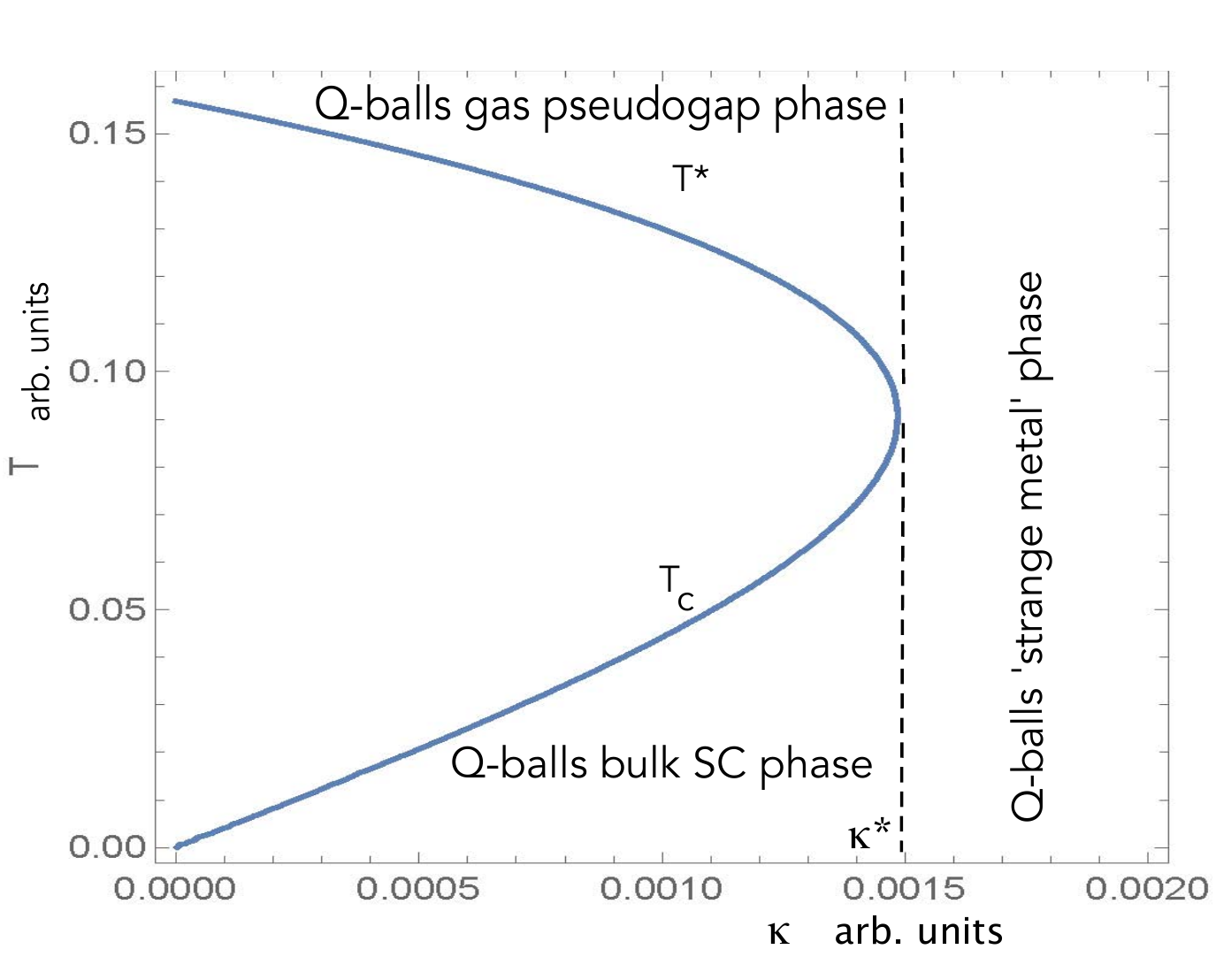}}
\caption{The phase diagram that follows from Eq. (\ref{self2}), where $\kappa\equiv c \dfrac{4g\nu \varepsilon_0}{3}$, see text.}
\label{T}
\end{figure}

\noindent It is obvious from Fig. \ref{triple} that: 1) at weak couplings the PG phase terminates at temperatures $T^*$ that are much higher than the temperatures $T_c$ of bulk superconducting transition; 2) there is some limiting coupling strength, at which $T^*$ touches $T_c$; 3) at even stronger couplings the expression on the l.h.s of Eq. (\ref{self1}) never touches zero at its minimum, but always crosses zero at two different values of $M/\Omega$, of which one approaches limit $M/\Omega=1$ of zero superconducting density, and the opposite one goes to 'infinity'. It is also noticeable from Fig. \ref{triple}, that local minima of  $\tilde{U}_{eff}$, that obey Eq. (\ref{self1}) for the different coupling strengths, are located nearly at one and the same coordinate along the $M/\Omega$ axis, i.e. for the fixed ratio: $M/\Omega=2$. Using this fact, one obtains the following approximate cubic equation, that provides the $T^*(\kappa)$ and $T_c(\kappa)$ dependences:

\begin{eqnarray}
 (\mu _0 ^2-\Omega^2) -c \dfrac{4g\nu \varepsilon_0}{3\Omega}=0; c=\left(\dfrac{\Omega}{M}\right)^2 I\left(\dfrac{M}{\Omega}\right)_{\frac{M}{\Omega}=2}\approx 0.01 \,. \label{self2}
\end{eqnarray}
\noindent The value of  $\kappa\equiv c \dfrac{4g\nu \varepsilon_0}{3}$, at which $T^*$ meets $T_c$, and respective temperature $T_0$ are:

\begin{eqnarray}
 \kappa^*=\dfrac{2\mu_0^3}{3^{3/2}}\,;\; T_c=T^*=T_0=\frac{\mu_0}{2\pi \sqrt{3}} \label{cross}
\end{eqnarray}
\noindent  The phase diagram that follows from Eq. (\ref{self2}) is plotted in Fig. \ref{T}.  To the right from the $T(\kappa)$ curve, i.e. for $\kappa > \kappa^*$, the 'PG' (PG) and superconducting phases are not divided, the Q-balls possess finite radii and $M/\Omega\approx 1$, according to the coordinates of the 'vertical' contours in Fig. \ref{triple} b), hence, the superconducting density approaches zero: $g_0=\sqrt{2M(M-\Omega)}\rightarrow 0$,  and superconducting transition acquires percolative character between chains of  the Q-balls connected with the Josephson links. This picture will be considered elsewhere.   

\section{The $T_c$ vs superconducting density $n_s$: the Uemura plot}
Obtained above solutions of the Q-ball self-consistency equation (\ref{self2}) and Eliashberg equation (\ref{Mg0}), it is possible to calculate the density of the superconducting condensate inside the Q-balls represented by diagonal value of  the Gor'kov Green's function $F$:

\begin{figure}[h!!]
{\includegraphics[width=0.55\linewidth]{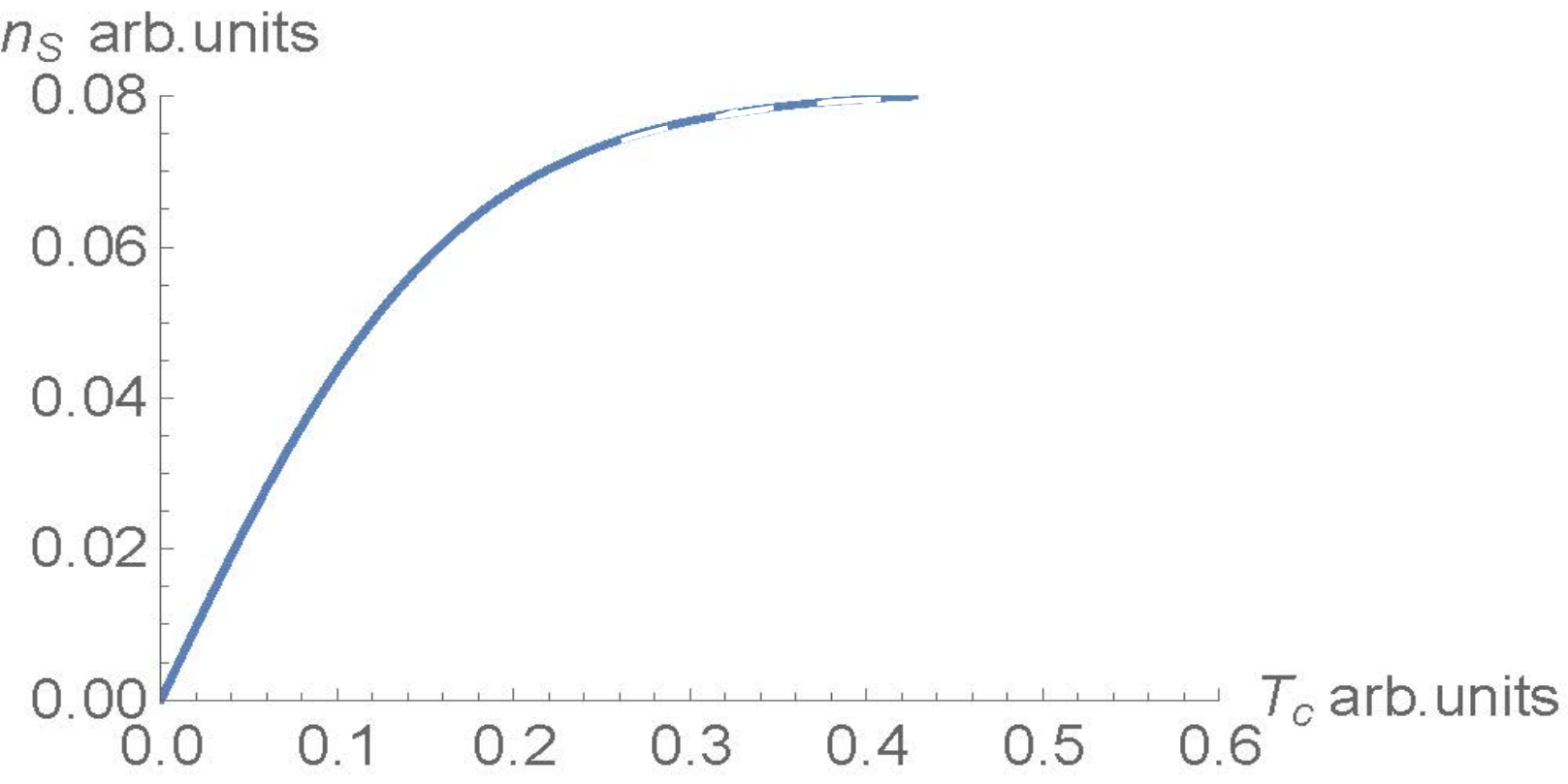}}
\caption{Calculated dependence of superconducting density $n_s$ as function of superconducting transition temperature $T_c$, see Eq. (\ref{ns1}). The dashed part of the curve overshoots the region $\kappa\ll\kappa^*$ on $T_c$ curve in Fig. \ref{T}, where validity of  linear approximation on temperature  $T_c$ is justified for the gap-function $g_0\sim T_c$ used in Eq. (\ref{ns1}), that generates the present plot, see text.}
\label{U}
\end{figure}

\begin{eqnarray}
 n_s = F(\vec{r},\vec{r}, \tau=0)=T\sum_{p,\omega} F_p(\omega)=\displaystyle\sum_p\frac{\sqrt{g_0^2-\varepsilon_p^2}}{2g_0}\tanh\frac{g_0}{2T}\,,
 \label{ns}
\end{eqnarray}
\noindent where in the last step an expression for the self-energy $\Sigma_2$ from Eq. (\ref{freeint}) was used. One has to take into account an expression for the density of the fermionic 'nested' states Eq. (\ref{nunest}) in order to accomplish summation over momentum space in Eq. (\ref{ns}), thus leading to the final result:

\begin{eqnarray}
 n_s = \displaystyle 2 {\int_0}^{g_0}\frac{\nu}{2g_0}\tanh\frac{g_0}{2T}\Theta(\varepsilon_0-\varepsilon)\sqrt{g_0^2-\varepsilon^2}\;d\varepsilon\approx\frac{\nu\varepsilon_0}{2}\tanh\frac{g_0}{2T}\tanh\frac{\pi g_0}{4\varepsilon_0} \,,
 \label{ns}
\end{eqnarray}
\noindent where the last factor extrapolates between the two cases $g_0<\varepsilon_0$ and $g_0>\varepsilon_0$. Now it is straightforward to substitute in Eq. (\ref{ns}) expression for $g_0$ from the self-consistency equation Eq. (\ref{Mg0}), and then use approximate relation $M=2\Omega$, valid for the linear region $\kappa\ll\kappa^*$ of superconducting transition points on $T_c$ curve in Fig. \ref{T} :

\begin{eqnarray}
 g_0=\sqrt{2M(M-\Omega)}\approx 2\Omega_c\,;\; n_s = \frac{\nu\varepsilon_0}{2}\tanh\frac{2\Omega_c}{2T_c}\tanh\frac{\pi \Omega_c}{2\varepsilon_0}\approx \frac{\nu\varepsilon_0}{2}\tanh\frac{\pi^2 T_c}{\varepsilon_0} \,,
 \label{ns1}
\end{eqnarray}
\noindent where one takes into account $\Omega_c\equiv 2\pi T_c $, leading to $\tanh\frac{2\Omega_c}{2T_c}=\tanh2\pi\approx 1$, see Fig. \ref{U}. Expression (\ref{ns1}) is remarkable: in the limit of  relatively small transition temperatures $T_c\ll \varepsilon_0/\pi^2$ it reproduces linear dependence of superconducting transition temperature $T_c$ on the density $n_s$ of the local-pair Bose-condensate in the Q-balls with the radius approaching infinity (bulk superconductivity transition):
 
\begin{eqnarray}
n_s \approx  \frac{\pi^2\nu}{2}T_c\,.  \label{ns2}
\end{eqnarray}

\noindent Here $\nu$  is the density of  fermionic 'nested' states (e.g. in the antinodal regions of cuprates fermi-surface) and may explain qualitatively the linear  dependence \cite{Uemura} on superconducting density of the superconducting transition temperatures $T_c$ in high-Tc superconducting compounds  found experimentally.

\section{Q-balls size}

\begin{figure}[h!!]
{\includegraphics[width=0.55\linewidth]{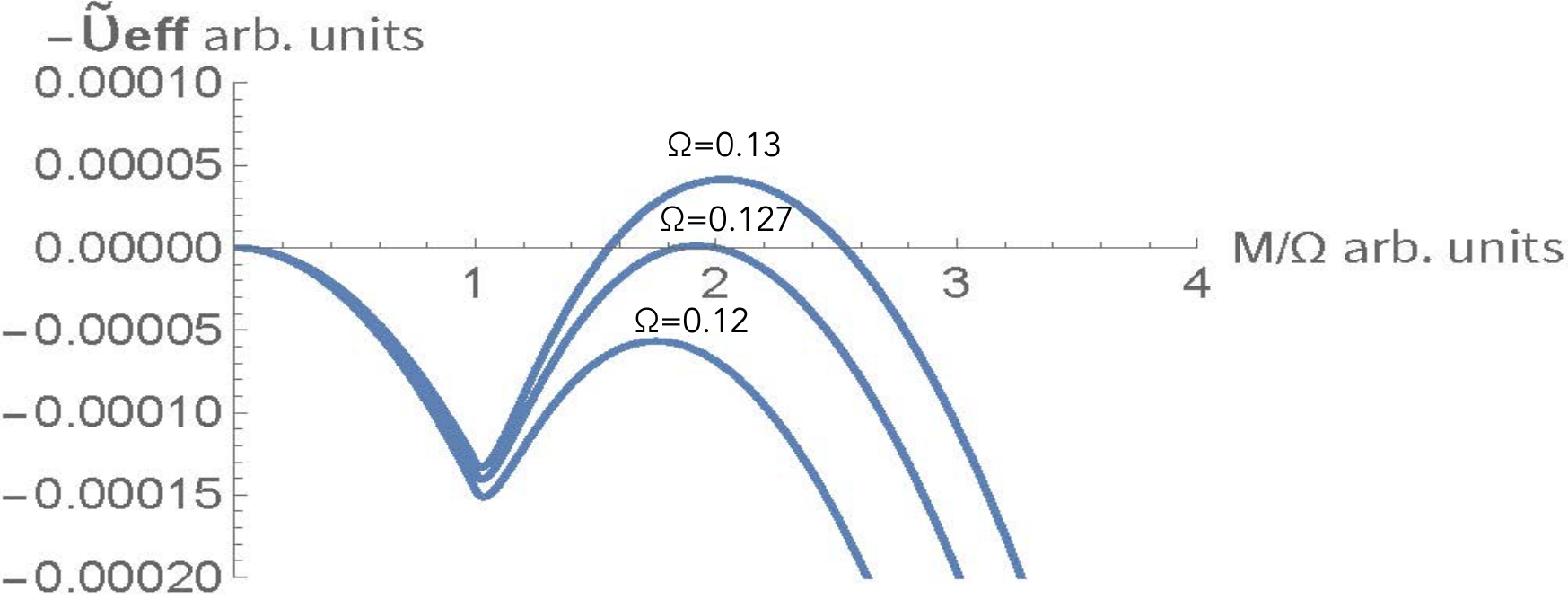}}
\caption{The plots of the 'reduced' potential energy $-\tilde{U}_{eff}$ for different values of  Matsubara frequency $\Omega=2\pi T$ at fixed coupling strength value $\kappa$ corresponding to contour curves in Fig.\ref{triple} with the same values as indicated in the caption. }
\label{Ueff}
\end{figure}

It is possible to understand relation between the Q-balls radii $R$ and the contour plots presented in Fig. \ref{triple} by investigating a complete coordinate dependent equation (\ref{selfr}) for the Q-ball field $M$, that minimises Euclidean action. Namely, using definition of  $\tilde{U}_{eff}$ in Eq. (\ref{self1}) and representation of Laplacian operator in spherically symmetric case one rewrites Eq. (\ref{selfr}) in the equivalent form:  

\begin{eqnarray}
\dfrac{d ^2M}{d r^2}=-\dfrac{2}{r}\dfrac{d M}{d r}-\dfrac{d\{- \tilde{U}_{eff}\}}{s^2d M}\,,  \label{Newt2}
\end{eqnarray}
\noindent that formally coincides with Newtonian equation of motion for a particle of unit mass in viscous environment moving in the potential 
$-\tilde{U}_{eff}$ , where radius $r$ plays the role of  'time' and modulus of fluctuation $M$ plays the role of 'coordinate', compare \cite{Coleman}. Neglecting 'damping' at large enough $r$, one finds an 'integral of motion':
\begin{eqnarray}
\dfrac{1}{2}\left\{\dfrac{d M}{d r}\right\}^2- \tilde{U}_{eff} =\tilde{E}=0\,,  \label{NwtI}
\end{eqnarray}
\noindent The integral of 'motion' $\tilde{E}$ is chosen to be zero taking into account finiteness of the Q-balls: $M(r>>R)=0$. Finally, 
$\dot{M}^2/2$ plays the role of 'kinetic energy'. Then, consider the plots of the potential  $-\tilde{U}_{eff}$ obtained using expressions Eq. (\ref{UFO1}), (\ref{self1}), see Figure \ref{Ueff}. It is straightforward to conclude from the conservation law (\ref{NwtI}) and Fig. \ref{Ueff} that coordinate $M$ of the 'particle' would take nearly 'infinite time' ($R\rightarrow \infty$) to reach point $M=0$ when it starts close to the top of the potential at 'initial time' ($r=0$), that happens when the maximum of $-\tilde{U}_{eff}$ touches axis $M/\Omega$. On the other hand, when $-\tilde{U}_{eff}$ crosses axis $M/\Omega$ at 'initial time' : $-\tilde{U}_{eff}(M(r=0))> 0$ it will take finite 'time' $R$ to reach point $M(R)=0$. Finally, when $-\tilde{U}_{eff}$ never crosses the axis $M/\Omega$ at any finite initial 'time' $r=0$ the finite time travel is not possible, i.e. no Q-ball solution exists for the case of the lowest curve $\Omega=0.12$ in Fig. \ref{Ueff}.

\begin{figure}[h!!]
{\includegraphics[width=0.55\linewidth]{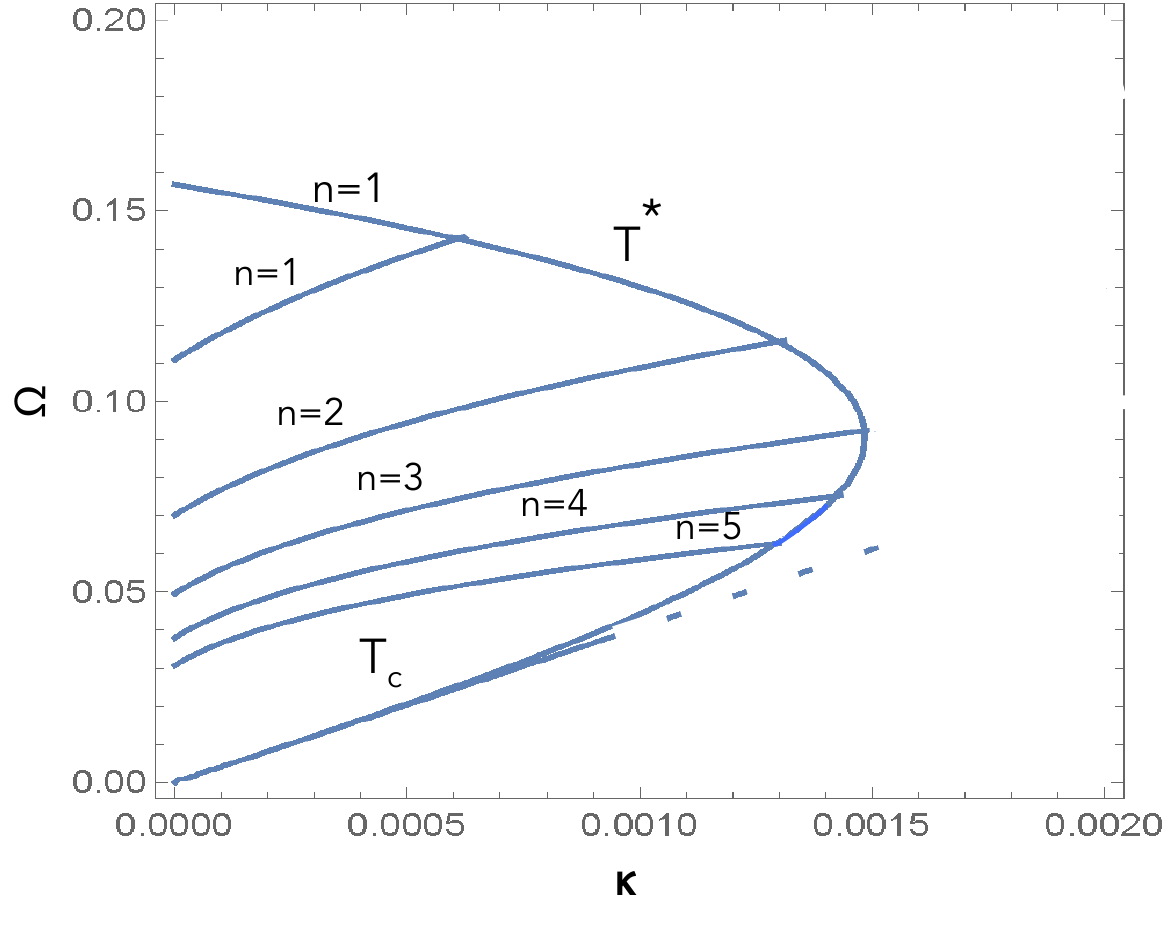}}
\caption{Solutions of Q-ball self-consistency equation (\ref{self}) approximated by Eq. (\ref{cubic1}) in the phase-space plane: coupling $\kappa$ -- frequency $\Omega$. Discrete lines marked with several integer values $n=1,2,3,4,5$ contain points in the phase-space plane, in which oscillations of the modulus $M$ of  spin/charge density occur inside the Q-balls. $\Omega = {\kappa}/{\mu _0 ^2}$ on the straight line labeled with symbol $T_c$, see text.}
\label{2b}
\end{figure}

\noindent  Finally, in order to distinguish different behaviours of the system when radius of the Q-ball becomes infinite at $T= T_c$ and $T=T^*$ it is important to check the sign of the Q-balls potential energy $U_{eff}$ given by Eq. (\ref{UFO1}), (\ref{UFOI}) in the two temperature intervals. It is most simple to check using approximate Eq. (\ref{self2}), from which it readily follows, that $U_{eff}<0$ when
$T\leq T_c$, hence bulk superconductivity at $R\rightarrow \infty $ takes place, while $U_{eff}>0$ when $T_c\leq T\leq T^*$, and therefore, the probability of the Q-ball with $R\rightarrow \infty $ goes to zero, hence, the Q-balls phase inside the loop $T_c(\kappa) -T^*(\kappa)$ in Fig. \ref{T} is possible only for integer $n=2,3,...$ in Eq. (\ref{self2}), thus leading to the temperature range $T_c\leq T\leq T^*/n $, see Fig. \ref{2b} and detailed derivation below.

It is possible to solve a complete coordinate dependent equation (\ref{selfr}) for the Q-ball field $M$, that minimises Euclidean action in a particular, but still quite general case for the contour plots presented in Fig. \ref{triple}. Namely, using found above analytical expression (\ref{UFOI}) for effective energy $U_{eff}$ and substituting it into coordinate dependent equation Eq. (\ref{selfr}) one obtains the following equation:

\begin{eqnarray}
-s^2\Delta M+(\mu_0^2-\Omega^2)M -\Theta(M-\Omega)\gamma\left[\dfrac{M\sqrt{2M(M-\Omega)}}{\Omega^2+8M(M-\Omega)}\right]\tanh\dfrac{g_0}{2T}\tanh\dfrac{g_0}{\varepsilon_0}=0\,,  \label{Iself}\\
g_0\equiv \sqrt{2M(M-\Omega)}\,;\;\gamma\equiv\dfrac{4g\nu \varepsilon_0}{3}\,. \label{Iselfr}
\end{eqnarray}
\noindent The analytic solutions of Eq. (\ref{Iself}) in the cases of  Q-balls with finite $R$, that correspond e.g. to the curve $\Omega=0.13$ in Fig. \ref{Ueff} could be found as follows. First, consider the case: $\Omega\ll g_0\ll \varepsilon_0$,  and substitute the last term in (\ref{Iself}) with approximate expression linear in $M$ :

\begin{eqnarray}
-s^2\Delta M+(\mu_0^2-\Omega^2)M -\Theta(M-\Omega)\dfrac{\gamma}{4\varepsilon_0}M=0\,.  \label{selflin}
\end{eqnarray}
\noindent Making then the usual substitution \cite{Flug}:

\begin{eqnarray}
M=\frac{\chi}{r}\,,\; M(0)< \infty\,,\label{chic}
\end{eqnarray}
\noindent one finds: 

\begin{eqnarray}
 &-s^2\ddot{\chi}+(\mu_0^2-\Omega^2){\chi}-\Theta(M-\Omega)\tilde{\gamma}{\chi}=0\,,  \label{selfch}\\
&\tilde{\gamma}\equiv\dfrac{\gamma}{4\varepsilon_0}=\dfrac{g\nu}{3}\,,
\end{eqnarray}
\noindent where ${\ddot{\chi}}\equiv d^2\chi/d r^2$. Next, one solves Eq. (\ref{selfch}) in the intervals $0<M<\Omega$ and $M>\Omega$ under the  continuity condition for the function $\chi(r)$ and its first derivative at the point $r=R$, that connects the two corresponding intervals of the spherical coordinate $r$: $\{0,R\}$ , $\{R, \infty\}$, where $R$ is the Q-ball radius. Then, when e.g. the case $M(R)=\Omega$ is chosen for definiteness, that corresponds to $Q=\frac{4\pi}{3}R^3\Omega^3$ according to Eq. (\ref{Q}), the result is as follows:

\begin{align}
M(r)\equiv\dfrac{\chi(r)}{r}=\begin{cases}\dfrac{ \sin{kr}}{r}\dfrac{\Omega R}{\sin{kR}} ;\;0<r\leq R;\vspace{0.2cm}\\
\dfrac{\Omega R}{r}\exp\{\lambda (R-r)\};\;R<r<\infty;\end{cases}
\label{chisol}
\end{align}
\noindent where:
\begin{eqnarray}
k=\frac{1}{s}\sqrt{\frac{g\nu}{3}+\Omega^2-\mu_0^2}\,;\;\lambda=\frac{1}{s}\sqrt{\mu_0^2-\Omega^2}\,; R=\frac{1}{k}\tan^{-1}{\frac{k}{\lambda}}\,. \label{alls}
\end{eqnarray}
\noindent Hence, the Q-balls of finite radii and energy do exist in the chosen limit $M\geq\Omega$, provided that the temperature (i.e. $\Omega$) belongs to the interval: $\sqrt{\mu_0^2-{g\nu}/{3}}<\Omega<\mu_0$.

Another analytic solution is possible to obtain for Eq. (\ref{self}), when the modulus $M$ in the definition Eq. (\ref{step}) is augmented with the $\tau$ dependence inside the Q-balls: $M(\tau,{\bf{r}})\rightarrow e^{-i\Omega\tau}M(\tau)$. In this case Eq. (\ref{self}) can be rewritten in the form of the 'energy conservation' relation for a one-dimensional motion of a 'particle' with coordinate $M(\tau)$ in the 'reduced' potential energy $\tilde{U}_{eff}(M)=U_{eff}(M)-\Omega^2M^2$:

\begin{equation} 
0=\dot{M}^2+\left(\mu _0 ^2-\Omega^2\right)M^2-\dfrac{4g\nu \varepsilon_0\Omega}{3}I\left(\dfrac{M}{\Omega}\right)\,.
\label{proenergy1}
\end{equation}
\noindent 
Then, equation (\ref{proenergy1}) possesses solution $M=const$, when minimum of  the 'reduced' potential energy $\tilde{U}_{eff}$ touches zero. It is easy to check that condition for the minimum $\partial \tilde{U}_{eff}(M)/\partial M=0$ reduces to an equation for the 'dimensionless' variable $M/\Omega$. One may use an approximate analytical expression for the integral in Eq. (\ref{UFOI}), that  leads to the following expression instead of (\ref{proenergy1}): 
\begin{eqnarray}
&0=\dot{M}^2 +\left(\mu _0 ^2-\Omega^2\right)M^2-M^2\dfrac{g\nu \varepsilon_0}{\Omega}f\left(\dfrac{M}{\Omega}\right)\,,\nonumber\\
&f\left(\dfrac{M}{\Omega}\right)\equiv \left\{  \dfrac{1}{3\sqrt{2}}\left(\dfrac{\Omega}{M}\right)^2  \left(\sqrt{\dfrac{M}{\Omega}\left(\dfrac{M}{\Omega}-1\right)}+ \ln\left\{\sqrt{\dfrac{M}{\Omega}}+\sqrt{\dfrac{M}{\Omega}-1}\right\}\right)\right\} \,.
\label{procubic}
\end{eqnarray}
\noindent  Minimising numerically function $f(z=M/\Omega)$ in the curly braces in the r.h.s of Eq. (\ref{procubic}) with respect to $M/\Omega$ one finds its value at the minimum: $f(z_0)=0.162$, that is reached at $z_0=M/\Omega=1.375$. Substituting these result  back into Eq. (\ref{procubic}) one finds the following   periodic in Matsubara time solutions of Eq. (\ref{procubic})  at discrete values of $\Omega_n=2\pi nT$:

\begin{align}
&\dfrac{M(\tau)}{\Omega}=\dfrac{z_0^2-c^2}{z_0+c\cos{\Omega_n \tau}}\,, c^2\equiv\dfrac{1}{\gamma}\left(1-\dfrac{\mu _0 ^2 \Omega-\Omega^3}{\kappa}\right)\,,\;\nonumber\\
&\gamma\equiv-\dfrac{f ^{''}(z_0)}{2f(z_0)}\approx 2.47\;,z_0\approx 1.38\,,\kappa=g\nu\varepsilon_0f(z_0) \label{oscill}
\end{align}
\noindent provided $\Omega_n$ obeys the following equation:

\begin{align}
\mu _0 ^2- \Omega^2-\dfrac{\kappa}{\Omega}\left(1-\gamma z_0^2\right)=\Omega_n^2\,,\;n=1,2, ...
\label{cubic1}
\end{align}
\noindent It is remarkable, that Matsubara time periodicity of bosonic semiclassical spin/charge density field amplitude $M(\tau)$ imposes quantisation of its oscillations around the static value $M=z_0\Omega$ i.e. around the point of the local minimum of Q-ball potential energy $\tilde{U}_{eff}(M)$, see Fig. \ref{2a}. The corresponding discrete lines for several integer values $n=1,2,3,4,5$ in the phase-space plane $\{\kappa,T\}$ are plotted in Fig. \ref{2b}.

 \noindent Finally, the Q-ball minimum of potential $U_{eff}$ touches zero, see Fig. \ref{2a}, when $\Omega$ obeys the following equation found by means of minimising and equating to zero of an approximate expression for $U_{eff}(M)$ provided by Eq. (\ref{procubic}) : $\Omega = {\kappa}/{\mu _0 ^2}$, which is plotted in Fig. \ref{2b} as straight line labeled with symbol $T_c$. On this line Q-ball volume becomes infinite according to self-consistency Eq. (\ref{self}), hence, bulk superconductivity transition must occur. 
 
 \section{The Q-balls Gas Thermodynamics in PG phase} 
Now, to explore different thermodynamic characteristics of the Q-balls in the PG phase,  we substitute into self-consistency equation of Q-ball emergence, expressed by Eq. (\ref{self}), an expression for effective potential due to superconducting fluctuations, derived above in Eq. (\ref{UFO1}), that close to $T^*$ acquires the form:

\begin{eqnarray}
 \Omega^2=\mu _0 ^2 -\dfrac{16g\nu M^{5/2}( M-\Omega)^{5/2}}{15\sqrt{2}T\Omega^2M^2}\;,\quad \mu^2_0/g\nu\ll1\;,
  \label{selfc}
\end{eqnarray}
\noindent where the last inequality we call condition of 'strong spin-fermion coupling'.
This section may be divided by subheadings. It should provide a concise and precise description of the experimental results, their interpretation as well as the experimental conclusions that can be drawn.

\subsection{$T_n^*$ and PG Phase}
Now, we solve self-consistency equation Eq. (\ref{selfc}) in the vicinity of the 1-st order phase transition temperature T$^*$ into pseudo gap phase \cite{Mukhin}: 
\begin{eqnarray}
M=\Omega\left(1+ \left( \dfrac{T^*_n-T}{\mu_0}\right)^{\frac{2}{5}}\left(\dfrac{15\mu^2 _0}{4\sqrt{2}g\nu}\right)^{\frac{2}{5}}\right),\quad T^*_n=\dfrac{\mu_0}{2\pi n}\;,
 \label{Mstar}
\end{eqnarray}
 \noindent where $n=1,2,...$. The highest value of T$^*_n$ corresponds to $n=1$.
 Substituting Eq. (\ref{Mstar}) into Eq. (\ref{Mg0}) one finds temperature dependence of the 'superconducting PG' :
 
 \begin{equation}
g_0^2= \left( T^*_n-T\right)^{\frac{2}{5}}{\Omega}^2\left(\dfrac{15\mu _0}{g\nu}\right)^{\frac{2}{5}}.
\label{g0star}
\end{equation}

 \begin{figure}[h!!]
\subfloat[a]{\includegraphics[width=0.45\linewidth]{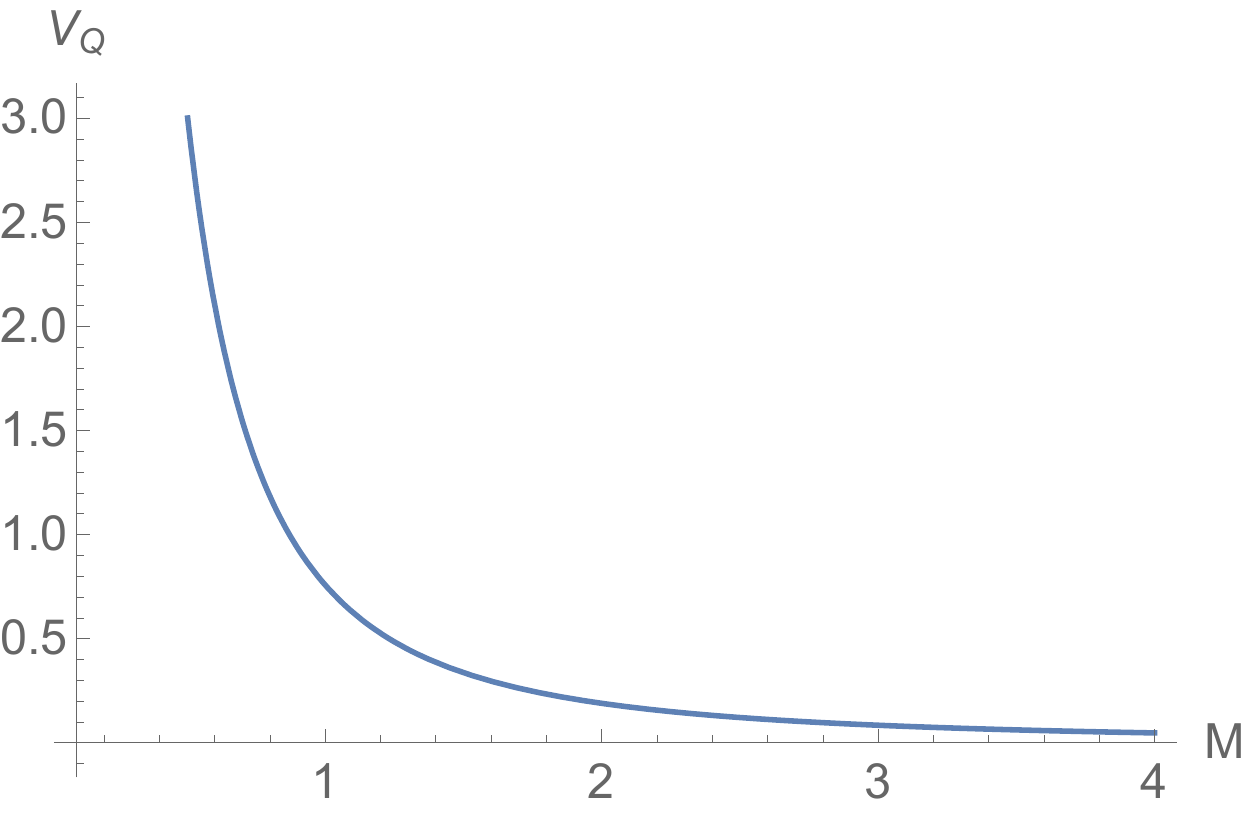}}\hspace{0.2cm}
\subfloat[b]{\includegraphics[width=0.45\linewidth]{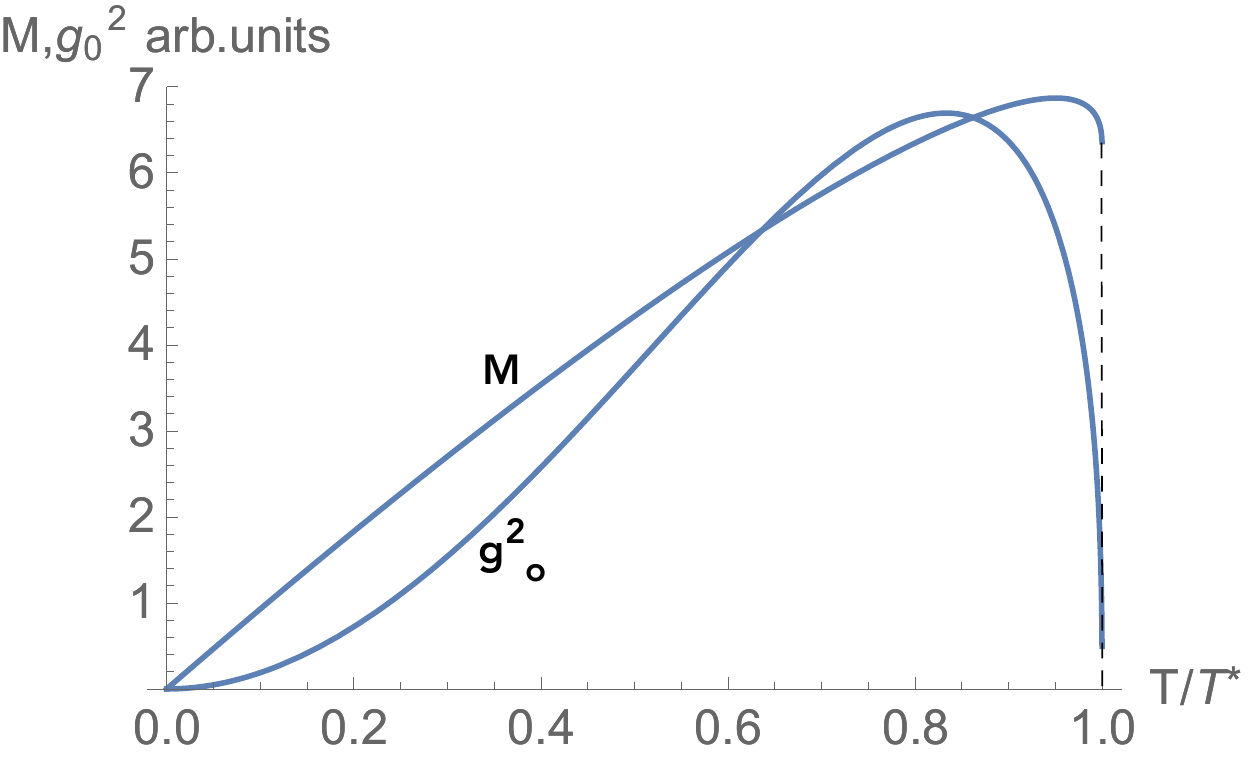}}
\caption{{\bf a}) Q-ball volume $V_Q$ as function of the spin-density wave amplitude $M$ ; {\bf b}) temperature dependence for the SDW/CDW amplitude $M$ and 'superconducting condensate density' $g_0^2$ inside a Q-ball in the units of  $T^*=\mu_0/2\pi n$ at fixed integer $n$.}
\label{4}
\end{figure}
\noindent It is remarkable that transitions at T$^*_n$ are of different types for semiclassical spin-density fluctuations and for superconducting condensate inside a Q-ball. While for the spin fluctuations amplitude  the transition is of the 1$^{st}$ order type , since it emerges at T$^*_n$ already with finite value $M\approx \mu_0$ , it is of the 2$^{nd}$ order type with respect to the superconducting fluctuation amplitude, as is manifested by Eq. (\ref{g0star}).  Hence, at T$^*_n$ at each $n$ there emerges a branch of Q-balls of a finite spin-density (charge-density) wave amplitude with zero superconducting condensate inside, the latter increases gradually and passes maximum when temperature lowers well below T$^*_n$. Simultaneously, substituting Eq. (\ref{Mstar}) into Q-ball volume formula (\ref{Q}) one finds:

\begin{eqnarray}
V_Q=\dfrac{Q}{\Omega M^2}=\dfrac{Q}{\Omega^3}\left\{1+ \left( \dfrac{T^*_n-T}{\mu_0}\right)^{\frac{2}{5}}\left(\dfrac{15\mu^2 _0}{4\sqrt{2}g\nu}\right)^{\frac{2}{5}}\right\}^{-2},\quad V_Q(T^*_n)=\dfrac{Q}{\mu_0^3}\propto Q\xi_M^3\;. \label{VT}
\end{eqnarray}
 \noindent Hence, at T$^*_n$ there emerges $n$-th branch of Q-balls of a volume proportional to spin fluctuations correlation volume $\sim\xi_M^3$ (modulo $Q$, considered below), where $\xi_M\sim 1/\sqrt{\mu_0}$ is (magnetic- /charge ordering) correlation length. The dependences signified by Eqs. (\ref{Q}) and (\ref{Mstar}), (\ref{g0star}) are plotted in Fig. \ref{4}. As long as $g_0$ characterizes superconducting gap in the fermionic spectrum, according to Eq. (\ref{freeint}) and also Fig. \ref{1},
 one concludes that Fig. \ref{4} provides a 'portrait' of the PG phase as the "profiles" of the density wave amplitude and gap in the fermionic spectrum'. 
 
\subsection{Finite size constraints on the minimal 'Noether charge' and specific heat of  Q-ball 'gas'} 
 Next, we consider consequences of the effect of finite size of Q-balls on the superconducting fluctuations.  Namely, self-consistency condition for emergence of superconductivity  in Eq. (\ref{Mg0}) was obtained under a disregard of the finiteness of the volume $V_Q$ of Q-ball fluctuation Eq. (\ref{VT}). To allow for the latter, one may apply linearised Ginzburg-Landau (GL) equation \cite{aaa} for the superconducting order parameter $\Psi$ to a Q-ball of radius $R$ in the spherical coordinates:

\begin{eqnarray}
-\dfrac{\hbar^2}{4m}\ddot{\chi}={bg_0^2}\chi\;; \quad\Psi(\rho)=\dfrac{C\chi(\rho)}{\rho}\;;\quad \Psi(R)=0, \label{GL}
\end{eqnarray} 
\noindent where $g_0^2$ from Eq. (\ref{Mg0}) substitutes GL parameter $a=\alpha \cdot(T_c-T)/T_c$ modulo dimensionfull constant $b$ of GL free energy functional \cite{aaa}. Then it is straightforward to deduce from a simple solution of Eq. (\ref{GL}):
\begin{eqnarray}
{\chi}\propto \sin(k_n\rho)\;;\quad Rk_n=\pi n,\;;\quad n=1,2,...,
\label{chik}
\end{eqnarray}
\noindent that in order for Eqs. (\ref{GL}) would possess solution (\ref{chik}) with the eigenvalue ${bg_0^2}$, the smallest radius $R_m$ of a Q-ball and corresponding volume $V_{Q_m}$ should obey the following conditions:

\begin{eqnarray}
&\dfrac{\hbar^2}{4m}\left(\dfrac{\pi}{R_m}\right)^2\leq {bg_0^2}\;, \rightarrow V^{1/3}_{Q}\geq V^{1/3}_{Q_m}=\left(\dfrac{Q_m}{\Omega M^2}\right)^{1/3}\equiv R_m= \pi\sqrt{\dfrac{\hbar^2}{4mbg_0^2}}\;. \label{Qm}
\end{eqnarray}

\begin{figure}[h!!]
\centerline{\includegraphics[width = 0.45\linewidth]{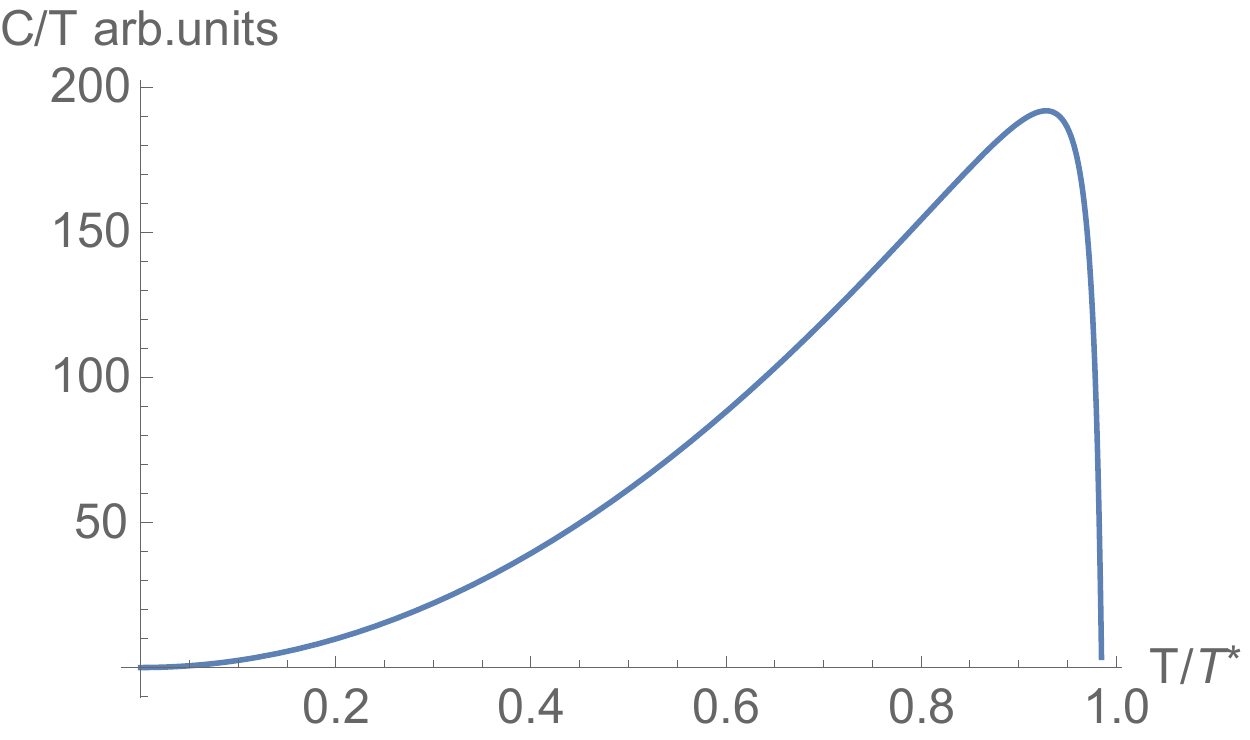}}
\caption{ Q-ball 'gas' contribution to specific heat $C$ as function of temperature in units of $T^*$.}
\label{3}
\end{figure}
 \noindent Now one can calculate Q-balls contribution to specific heat using for their entropy the thermodynamic expression for the Boltzmann 'gas' \cite{LL5} :  
 
\begin{eqnarray}
{\cal{ S}}_Q=\sum_{Q,n}G_{Q,n}\bar{n_{Q,n}}\ln{\dfrac{e}{\bar{n_{Q,n}}}};\quad \bar{n_{Q,n}}=\exp{\left\{-\dfrac{E_{Q,n}}{k_BT}\right\}}=\exp{\left\{-\dfrac{2Q\Omega}{gk_BT}\right\}} \;,
 G_{Q,n}=\dfrac{V}{V_{Q}},
 \label{LL5}
\end{eqnarray}  
 \noindent where 'coordinates' $\{Q,n\}$ span the phase space of the Q-balls formed by the values of the 'Noether charge' $Q$ and discrete
 values of the Matsubara frequencies $\Omega\equiv 2\pi nT$, $n=1,2,...$.  The number of the different 'positions' of a $Q,n$-ball in real space is evaluated as  ${V}/{V_{Q,n}}$, where $V$ is the volume of the system. The Boltzmannian exponent in Eq. (\ref{LL5}) contains Q-ball energy expressed in Eq. (\ref{aQ}). Near T$^*_n$, using Eqs. (\ref{g0star}), (\ref{Qm}) we evaluate lower bound $Q_m$ in the summation over Q in Eq. (\ref{LL5}) and restrict the sum to the contribution of the $n$-th branch when $T\simeq T^*_n $, thus finding the following contribution of Q-balls to the entropy and specific heat of the system:
 
 \begin{eqnarray}
\dfrac{C_{V,n}}{T}=\dfrac{\partial{\cal{ S}}_{Q,n}}{\partial T}\propto V\dfrac{\Omega^3}{T}\left(1-\dfrac{C\Omega}{5g(T^*_n-T)^{8/5}}\right)\exp{\left\{-\dfrac{C2\pi n}{g(T^*_n-T)^{3/5}}\right\}},
 \label{Cstar}
\end{eqnarray}  
\noindent where constant $C$ has absorbed all the dimensionfull constants from Eqs. (\ref{g0star}), (\ref{Qm}), and it is assumed that argument of the exponential function is mach greater than 1 close to T$^*_n$, see Fig. \ref{3}. 
\section{Diamagnetic response of Q-ball gas}

\begin{figure}[h!!]
\centerline{\includegraphics[width=0.5\linewidth]{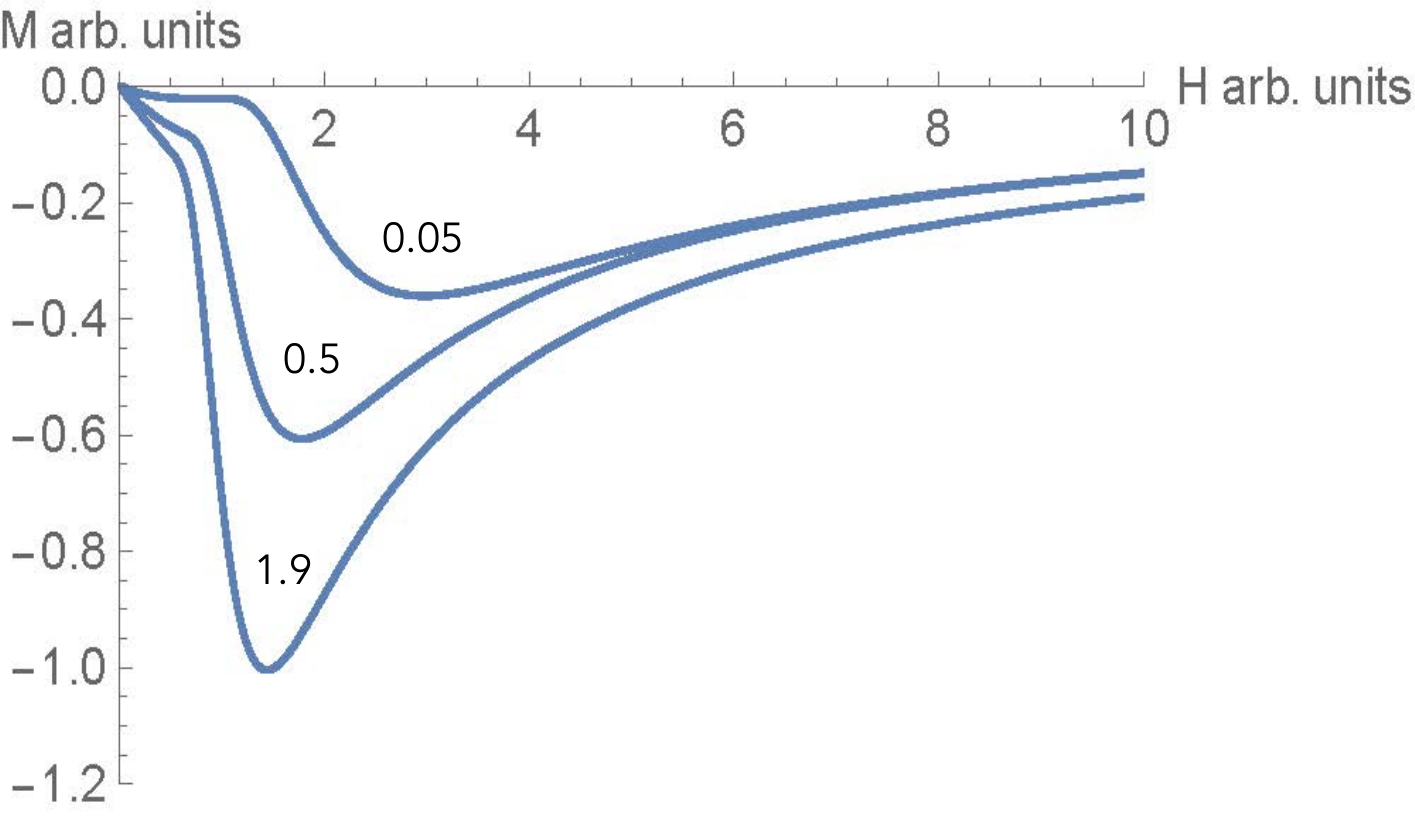}}
\caption{ Density of diamagnetic moment of the Q-balls gas in the PG phase $T^*(\kappa)<T<\mu_0/(2\pi )$ , curves 1-3 correspond to different values of temperature $\mu_0/(2\pi )-T$ indicated in arb. units, see Fig. \ref{triple} and Eqs. (\ref{Mstar}), (\ref{M1}), (\ref{M2}).}
\label{dia}
\end{figure}

It is straightforward to apply presented above picture of Q-ball gas in high-T$_c$ superconductors for description of experimentally discovered diamagnetic behaviour above Tc in cuprates \cite{li, zaanen}. 
Again, as in Eq. (\ref{LL5}) using the concept of the phase space of the Q-balls formed by the values of the 'Noether charge' $Q$ and discrete
 values of the Matsubara frequencies $\Omega_n\equiv 2\pi nT$, $n=1,2,...$, and counting the number of the different 'positions' of a Q-ball in the real space as  ${V}/{V_{Q,n}}$, where $V$ is the volume of the system and the Q-ball volume is determined using the 'charge' Q conservation law Eq. (\ref{Q}):

 \begin{eqnarray}
V_{Q,n}\equiv \dfrac{4\pi R^3}{3}= \dfrac{Q}{\Omega_n M^2}\; , \label{QQ}
\end{eqnarray}
\noindent one finds the following expression for the partition function of the Q-balls gas in the temperature range where it exists, $T^*(\kappa)<T<\mu_0/(2\pi n)$, see Eq. (\ref{self2}) and Fig. \ref{triple}:

\begin{eqnarray}
Z_Q=\sum_{Q,n}\frac{1}{N!}\left[\int_{Q_m}^{Q_H}dQ\frac{V}{V_{Q,n}}\exp\left\{-\left[\dfrac{2Q\Omega_n}{gT}-\dfrac{M_QH}{T}\right]\right\}\right]^N   \;,
  \label{MQ}
\end{eqnarray}  

\noindent
 The Q-ball energy in the first term of the Boltzmann's expression in the brackets in Eq. (\ref{MQ}), $E_Q/T$, is taken from the self-consistency Eq. (\ref{aQ}). The lower and upper bounds in the integral over $dQ$ are as follows. The smallest value of $Q=Q_m$ is obtained from Eq. (\ref{QQ}) for the Q-ball of the size $R_m$ bound from below by the Landau correlation length $\xi$, see Eq. (\ref{Qm}): 

\begin{eqnarray}
&{Q_m}={\Omega M^2}\dfrac{4\pi R_m^3}{3}\,,\;R_m=\xi\equiv \pi\sqrt{\dfrac{\hbar^2}{4mbg_0^2}}\;. \label{Qmm}
\end{eqnarray}
\noindent with $g_0$ defined by Eq. (\ref{Mg0}). The upper bound $Q_H$ in the integral in Eq. (\ref{MQ}) is obtained as follows:

\begin{eqnarray}
&{Q_H}={\Omega M^2}\dfrac{4\pi R_H^3}{3}\,,\;R_H=\dfrac{\delta_LH_c\sqrt{20}}{H}\,,\;\delta_L=\dfrac{\sqrt{mc^2}}{\sqrt{4\pi n_se^2}}\,, \label{QH}
\end{eqnarray}
\noindent where $R_H\ll \delta_L$ is the maximum radius of a small superconducting sphere \cite{LL9}, at which it remains  superconducting in magnetic field $H$, and $\delta_L$ is London penetration depth, $H_c$ is critical magnetic field of the bulk superconductor material,  $n_s$ is superconducting electrons density given in Eq. (\ref{ns2}), $m$ is electron mass, and $c$ is light velocity. 
\noindent The next term, $-M_QH/T$, in the Boltzmann's expression in the brackets in Eq. (\ref{MQ}) is the energy of diamagnetic moment $M_Q$ in magnetic field $H$:

\begin{eqnarray}
&{M_Q}=-\dfrac{R^5H}{30\delta_L^2}H=-\left(\dfrac{3Q}{4\pi M^2\Omega}\right)^{\frac{5}{3}}\dfrac{H^2}{30\delta_L^2}\,, \label{MQLL}
\end{eqnarray}
\noindent where $M_Q$ is projection of diamagnetic moment of  a Q-ball on the magnetic field direction $\vec{H}$. The Q-ball is regarded as a small superconducting sphere of radius $R\ll \delta_L$ possessing diamagnetic moment in magnetic field $H$ \cite{LL9}. In the last equality in Eq. (\ref{MQLL}) $R$ is  substituted via the expression $R=R(Q)$ obtained from the Q-ball 'charge' $Q$ conservation relation Eqs. (\ref{Q}), (\ref{QQ}). Composing altogether the above relations one finds the following expression for the free energy of the Q-ball gas:

\begin{eqnarray}
&F=-T\ln{Z_Q}\,,\; Z_Q=\displaystyle \sum_{n,N}\dfrac{G_n^N}{N!}\equiv \exp{G_n}\,,\label{FQ}\\
&G_n= \displaystyle\int_{Q_m}^{Q_H}dQ\dfrac{V\Omega_n M^2}{Q}\exp\left\{-\left[\dfrac{2Q\Omega_n}{gT}+\left(\dfrac{3Q}{4\pi M^2\Omega_n}\right)^{\frac{5}{3}}\dfrac{H^2}{30\delta_L^2T}\right]\right\}   \;,\label{GQ}\\
&Q_H=\dfrac{\delta_L^3H_c^3}{H^3}\dfrac{4\pi\Omega_nM^2 20^{\frac{3}{2}}}{3}  \label{QHF}
\end{eqnarray} 
\noindent In the highest temperature interval $T^*(\kappa)<T<\mu_0/(2\pi n)$ one takes integer $n=1$, see Eq. (\ref{self2}) and Fig. \ref{triple}a), and then for the free energy of the "hot" Q-balls gas and its density of diamagnetic moment $<M_Q>/V$ one finds: 

\begin{eqnarray}
&F=-TG_{n=1}\equiv -TG\,,\; <M_Q/V>=T\dfrac{\partial G}{V\partial H}\equiv- M_1-M_2\,,\label{Fn1}\\
&M_1=\dfrac{2H3^{5/3}}{30\delta_L^2(4\pi)^{5/3}( M^2\Omega)^{2/3}}\displaystyle\int_{Q_m}^{Q_H}dQ{Q}^{2/3}\exp\left\{-\left[\dfrac{2Q\Omega}{gT}+\left(\dfrac{3Q}{4\pi M^2\Omega}\right)^{\frac{5}{3}}\dfrac{H^2}{30\delta_L^2T}\right]\right\}\,,\label{M1}\\
&M_2=\dfrac{3\Omega M^2}{H}\exp\left\{-\left[\dfrac{2Q_H\Omega}{gT}+\left(\dfrac{3Q_H}{4\pi M^2\Omega}\right)^{\frac{5}{3}}\dfrac{H^2}{30\delta_L^2T}\right]\right\}\,,\label{M2}
\end{eqnarray}
\noindent where one has to substitute solution $M=M(\Omega)$ of the self-consistency Eq. (\ref{aQ}) using e.g. solutions from Eq. (\ref{Mstar}), or in the form of contour plots in Fig. \ref{triple}. This leads to the following dependence found numerically from Eqs. (\ref{M1}), (\ref{M2}) above, see Fig. \ref{dia}.

\section{Conclusions}
\label{sec: fin}
To summarise, a 'pairing glue' by exchange with coherent semiclassical fluctuations inside finite volume nontopological Euclidean solitons, Q-balls, is proposed as a mechanism of PG (PG) phase and high temperature superconductivity in high-$T_c$ cuprates.
It is demonstrated that Euclidean Q-balls of semiclassical spin-/charge density-wave fluctuations, that self-consistently support formation of local superconducting condensates, can emerge as 'smoking gun' of PG phase and high temperature superconductivity in strongly enough coupled repulsive Fermi systems with 'nested' regions of the Fermi surface with finite density of fermionic states. The proposed theory of pairing via  exchange with semiclassical fluctuations of finite amplitude at the local minimum of their potential energy inside the Q-balls differs from the standard Fr\"{o}hlich pairing mechanism via  exchange between fermions with incoherent bosons of infinitesimal amplitudes, e.g. phonons \cite{elis}, spin-waves \cite{Chubukov}, or polarons \cite{Bianconi}. Proposed here theory is simple enough,  so that it could provide basis for an analytically treatable calculations of spectral \cite{campi, caprara}, transport, thermal \cite{Taillefler} and electromagnetic \cite{li} properties of the high temperature superconductors in  PG and superconducting states. Besides, the superconducting transition may acquire percolative character due to Josephson tunneling between Q-balls forming infinite percolating clusters. This picture will be considered elsewhere and compared with the known properties of the 'strange metal' phase beyond the optimal doping \cite{zaanen}.  As a first step, it is demonstrated in the above  Sections, that presented theory of Q-balls formation may naturally explain the linear  dependence of   $T_c$ on superconducting density $n_s$ in high-Tc superconducting compounds  found experimentally \cite{Uemura}, as well as diamagnetism combined with Cooper pairing above Tc in cuprates \cite{li}. Prediction of a sharp maximum in specific heat temperature dependence in the vicinity of the 1-st order phase transition into Q-balls gas phase , that follows from proposed above theory is also presented.  It is also interesting to admit that obtained Q-ball solutions fall into the category of finite size thermodynamic time crystals, considered previously \cite{3,4,5,6}.

\acknowledgments{The author acknowledges useful discussions with Serguey Brazovskii, Jan Zaanen, Carlo Beenakker, Konstantin Efetov and Andrey Chubukov. This research was supported by the Ministry of Science and Higher Education of the Russian Federation in the framework of Increase Competitiveness Program of NUST MISiS Grant No. K2-2020-038}.

\end{document}